\documentclass[twocolumn]{aastex631}

\usepackage{graphicx}	
\usepackage{amsmath}	
\usepackage{amssymb}	
\usepackage{dsfont}
\usepackage{newtxtext,newtxmath}    
\usepackage{ulem}
\usepackage{soul}
\usepackage{hyperref}
\usepackage{macros_vdbosch} 

\DeclareRobustCommand{\VAN}[3]{#2}
\let\VANthebibliography\thebibliography
\def\thebibliography{\DeclareRobustCommand{\VAN}[3]{##3}\VANthebibliography}

\hypersetup{draft}

\begin{document}

\title[The Dipole Instability in Gravitational Systems]
      {The Dipole Instability in Gravitational $N$-body Systems: A Natural Explanation for Lopsidedness and Off-Centered Nuclei in Galaxies}

\author{Shashank Dattathri}
\affiliation{Department of Astronomy, Yale University, PO. Box 208101, New Haven, CT 06520-8101}

\author{Frank C. van den Bosch}
\affiliation{Department of Astronomy, Yale University, PO. Box 208101, New Haven, CT 06520-8101}

\author{Martin D. Weinberg}
\affiliation{Department of Astronomy, University of Massachusetts, Amherst, MA 01003-9305, USA}

\author{Uddipan Banik}
\affiliation{Department of Astrophysical Sciences, Princeton University, 112 Nassau Street, Princeton, NJ 08540, USA}
\affiliation{Institute for Advanced Study, Einstein Drive, Princeton, NJ 08540, USA}
\affiliation{Perimeter Institute for Theoretical Physics, 31 Caroline Street N., Waterloo, Ontario, N2L 2Y5, Canada}

\label{firstpage}

\begin{abstract}
  We explore the stability of isotropic, spherical, self-gravitating systems with a double-power law density profile. Systems with rapid transitions between the inner and outer slopes are shown to have an inflection in their isotropic distribution function (DF), where $\rmd f/\rmd E > 0$, thereby violating Antonov's stability criterion. Using high-resolution $N$-body simulations, we show that the resulting instability causes the growth of a rotating dipole (or $l=1$) mode. The inflection feature in the DF responds to the mode by promoting its growth, driving the instability. The growth of the dipole results in a torque that dislodges the original cusp from its central location, and sets it in motion throughout the central region. Once the mode goes non-linear, it saturates, together with the cusp, into a long-lived soliton (the $l=1$ equivalent of a bar in a disk galaxy), which maintains its sloshing motion through the center of the halo along a slowly precessing, elliptical orbit. Concurrently, the soliton traps increasingly more particles into libration, and the exchange of energy and angular momentum with these trapped particles works towards eroding the bump in the distribution function. We point out similarities between the dipole mode and the bump-on-tail instability in electrostatic plasmas, and highlight a potential connection with core stalling and dynamical buoyancy in systems with a cored density profile. Finally, we discuss the astrophysical implications in terms of lopsidedness and off-center nuclei in galaxies.
\end{abstract}



\section{Introduction}
\label{sec:intro}

Characterizing the detailed, dynamical structure of galaxies and their dark matter halos is an important goal of modern astrophysics. Often, relatively simple equilibrium models are used to describe these systems without addressing their stability.  However, it turns out that many self-gravitating, collisionless $N$-body systems have unstable equilibria. Hence, if the complex processes associated with galaxy formation try to establish such an equilibrium, it will develop a response which rapidly grows in time, driving the system towards a new, dynamical equilibrium. In addition, even stable systems will typically fluctuate around their equilibrium configurations. Understanding which systems are stable or unstable, and how they evolve or fluctuate, is important for understanding the observed morphological and kinematic diversity among galaxies and their structural details.

The stability of a gravitational system, which obeys both the collisionless Boltzmann equation (CBE) and the Poisson equation, can be analyzed using a modal analysis based on linear response theory. This approach has recently become more accessible due to the development of the public code {\tt LinearResponse.jl} \citep[][]{Petersen2024}. The slowly evolving linear solutions of the coupled CBE-Poisson equation can be written as a generalized eigenvalue equation (or alternatively, as a Volterra integral equation). By choosing an appropriate set of biorthogonal basis functions to represent the density and potential fields, this solution can be expressed in terms of the linear response matrix $\bM(\omega)$ \citep[e.g.,][]{Kalnajs1977, Fridman.Polyachenko.84, Palmer1987, Weinberg1991, Hamilton.Fouvry.24}, which is given by:

\begin{equation}
\label{eq:response_matrix}
    M_{\rm pq}(\omega) = \frac{(2\pi)^3}{4 \pi G} \sum_\mathbf{\ell} \int \rmd\mathbf{J} \frac{\mathbf{\ell} \cdot {\partial f}/{\partial \mathbf{J}}}{\omega - \mathbf{\ell}\cdot \mathbf{\Omega}} \Phi_{\mathbf{\ell}}^{(\rmp)*}(\mathbf{J}) \Phi_{\mathbf{\ell}}^{(\rmq)}(\mathbf{J})
\end{equation}
where $\mathbf{J}=(J_1,J_2,J_3)$ are the actions, $\mathbf{\Omega}=(\Omega_1,\Omega_2,\Omega_3)$ are their corresponding frequencies, $\mathbf{\ell}=(l_1,l_2,l_3)$ is a vector of integers, $f$ is the (unperturbed) distribution function (DF), and $\Phi_{\mathbf{\ell}}^{(q)}(\mathbf{J})$ is the Fourier transform in angle variables of the basis elements of the potential field.

The modes of the system are then given by the zeros of the dispersion relation 
\begin{equation}
\label{eq:dispersionrelation}
\mathcal{D}(\omega)={\rm det} \left(\bI - \bM(\omega) \right) = 0\,,
\end{equation}
where $\bI$ is the identity matrix. The modes are discrete in frequency space and are referred to as point modes or Landau modes, where the latter nomenclature has been borrowed from plasma physics\footnote{While it is common to refer to the point modes of gravitational systems as Landau modes, there are subtle differences between Landau (point) modes in self-gravitating systems and Landau modes in plasmas (see Section~\ref{ssec:landau}).}. The complex frequencies, $\omega$, describe self-similar solutions with time dependence $\exp(-i\omega t)$. Hence, systems that have Landau modes with a positive imaginary frequency are unstable. The most well-known example of such a Landau mode in a gravitational $N$-body system is the unstable Jeans mode of a homogeneous density distribution. Another example, which is particularly relevant for this paper, is the lopsided $l=1$ (or dipole) mode in spherical systems first identified by \citet{Weinberg1994} and later examined in more detail in \citet{Heggie2020}. Although in all these cases this $l=1$ mode was found to be weakly damped, in this paper we shall encounter the same mode but with a positive imaginary frequency, thus growing over time. 

An alternative to modal analysis is to set up the initial conditions (IC) of the system in equilibrium and evolve it forward in time using an $N$-body simulation. If the system evolves away from the ICs it may be deemed unstable. Although fairly straightforward in principle, in practice, one is faced with the challenge to demonstrate that any evolution reflects a true instability rather than some numerical artifact. This is best addressed by comparing the simulation results with predictions based on a linear mode analysis, which is the methodology that we adhere to in this paper. Simulations have the added benefit that they can be used trivially to study the system's evolution even after the mode goes non-linear, which is not possible with linear response theory.

The simplest realistic systems to consider for stability analysis are spherical systems with an isotropic distribution function, $f=f(E)$. Using the variational principle, \citet{Antonov1962} demonstrated that such systems are stable to non-radial modes provided that $\rmd f/\rmd E < 0$. Later, \citet{Doremus1971} showed that such systems are also stable to radial modes. Hence, spherical systems with an isotropic distribution function satisfying $\rmd f/\rmd E < 0$ are stable against \textit{all} perturbations. Several authors have extended this work to analyze spherical systems with anisotropy. For example, systems with a strong radial anisotropy are known to be unstable due to a radial orbit instability (ROI) that causes them to rapidly evolve into elongated, bar-like configurations. The ROI was first independently predicted by \citet{Antonov1973} and \citet{Henon.73}, and subsequently confirmed numerically by \citet{Polyachenko1981}, \citet{Merritt1985} and \citet{Barnes1986}. 

Note that it has only been proven that $\rmd f/\rmd E < 0$ is a {\it sufficient} condition for stability. It is not guaranteed to be a {\it necessary} condition; that is, it is not guaranteed that a system that violates this condition must be unstable. In fact, \citet{Goldstein.71}, \citet{Doremus.Feix.72}, and \citet{Henon.73} have all presented counterexamples of systems that are stable while violating the condition $\rmd f/\rmd E < 0$. However, all these systems are one-dimensional, with little relevance to astrophysical systems, or, as in the case of \citet{Henon.73}, only consisting of purely radial modes. Recently, a more relevant case was brought to light by \citet{Weinberg2023}, who showed that a spherically symmetric NFW \citep{Navarro.etal.1997} profile that is truncated in its outskirts can have an inflection in its DF where $\rmd f/\rmd E > 0$. He then showed, using both modal analysis and numerical simulations, that the system is unstable in that it develops a rapidly growing, oscillating dipole mode. Following a rapid, linear growth, the mode ultimately saturates into a long-lived soliton that sloshes back-and-forth through the center of the halo. Such modes could have potentially far-reaching implications for galaxies that reside in such halos, warranting a more detailed investigation. 

If local inflections, or bumps, in the distribution function of a gravitational $N$-body system can spawn instabilities, the natural question is when and where such features might be encountered in the DF. The inflection in the DF of the system studied by \citet{Weinberg2023} is a consequence of the fact that the density profile of the system was truncated at large radii. Although truncation is natural, given that each halo is embedded in a tidal field of some sort, not every truncation necessarily results in a bump, and it is possible to set up a truncated density profile that is stable without introducing a bump in the DF\footnote{For example, the well known family of \citet{King1966} models, originally introduced by \citet{Michie.63}, are truncated by simply setting the DF for an untruncated isothermal sphere to zero for $E > E_{\rm crit}$, where $E_{\rm crit}$ is the critical truncation energy, and then shifting the DF downward such that $f(E_{\rm crit})=0$. This method is often used to set up equilibrium, truncated density profiles for idealized numerical simulations \citep[][]{Drakos.etal.17, Chiang.etal.24}, or to describe systems that are tidally stripped \citep[][]{Drakos.etal.20}.}. However, truncation is not the only mechanism that can give rise to inflections in the DF. As pointed out by \citet{Weinberg2023}, a wide variety of natural processes associated with the hierarchical assembly of DM halos, such as the accretion and disruption of subhalos and incomplete mixing following a merger event, can introduce special features in the DF that break its monotonic energy dependence. Hence, the hierarchical formation of halos and galaxies may naturally lead to inflections, potentially triggering a dipole mode.

In this paper, we show that inflections in the isotropic DF $f(E)$ can be present in the {\it central} regions of spherical systems with a double power-law density profile, if the transition from the steep outer profile to the shallower inner profile is sufficiently rapid. Using high-resolution $N$-body simulations we demonstrate that, similar to Weinberg's truncated halo, such systems develop a rotating dipole mode in the central region which rapidly saturates into a long-lived soliton that sloshes back and forth through the system's core region, similar to a seiche in a lake or swimming pool. The mode dislodges the central cusp, which sloshes through the central region of the galaxy, in much the same way as a beach ball sloshes around in a swimming pool with a seiche. We study the resonant interactions between the dipole mode, the cusp, and individual particles and how this impacts the evolution of the system's DF. This elucidates that the lopsided $l=1$ instability studied here is the analog of the well known bump-on-tail instability in plasma physics, but for an inhomogeneous gravitational system. We argue that the physics that underlies the saturation of the mode is similar to what causes the saturation of spiral and bar modes in disk galaxies. In fact, as pointed out in \citet{Weinberg2023}, the long-lived $l=1$ soliton that forms as a result of this saturation is basically the $l=1$ equivalent of the $l=2$ bar in a disk.

This paper is organized as follows. Section~\ref{sec:abg} examines the distribution functions of isotropic spheres with a generalized double power-law density profile. We demonstrate that there is a large part of parameter space for which the distribution functions reveal an inflection (i.e., a `bump'). Section~\ref{sec:method} describes the $N$-body simulations that we use to probe the evolution of isotropic spheres with such a bump, as well as the methodology used to analyze these simulations. Section~\ref{sec:evolution} shows that the inflection indeed results in an $l=1$ instability and gives a detailed description of the growth and saturation of the mode. In Section~\ref{sec:dynamics} we study the detailed interaction between the $l=1$ mode and individual particles, which gives valuable insight into the physical mechanisms that lead to the growth of the dipole mode. Section~\ref{sec:discussion} puts our findings in perspective by pointing out similarities with the bump-on-tail instability in plasma physics, the bar instability in disk galaxies, and by highlighting how the mechanism that underlies all these instabilities also governs dynamical friction, core stalling, and dynamical buoyancy. We also discuss the potential implications of $l=1$ modes for the evolution of galaxies and conclude in Section~\ref{sec:conclusion}.


\section{Isotropic spheres with a double power-law density profile}
\label{sec:abg}
 
Galaxies and dark matter halos (to which we collectively refer as \textit{systems}) exhibit a diverse range of density profiles. However, a commonality shared by many is that their density profile resembles a double power-law, which can be characterized by the general $\abg$ profile, first introduced by \citet{Zhao1996}:
\begin{equation}
\label{eq:dens}
    \rho(r) = \rho_0 \, \left( \frac{r}{r_\rms} \right)^{-\gamma} \left[ 1 + \left( \frac{r}{r_\rms} \right)^{\alpha} \right]^{\frac{\gamma-\beta}{\alpha}}  \,.
\end{equation}
Here $r_\rms$ is the scale radius and $\rho_0$ is a normalization constant. This profile transitions from an outer power-law, $\rho \propto r^{-\beta}$, to an inner power-law, $\rho \propto r^{-\gamma}$, with $\alpha$ controlling the steepness of the transition. The subset of models with $\alpha=1$ and $\beta=4$ has been studied in detail by \citet{Dehnen1993} and \citet{Tremaine1994}, and includes the well-known \citet[][$\gamma=1$]{Hernquist1990} and \citet[][$\gamma=2$]{Jaffe1983} profiles. Other well-known examples of $\abg$-profiles are the NFW profile, which has $\abg=(1,3,1)$ and the \citet{Plummer1911} profile, which has $\abg=(2,5,0)$.
\begin{figure}
    \centering
    \includegraphics[width=\columnwidth]{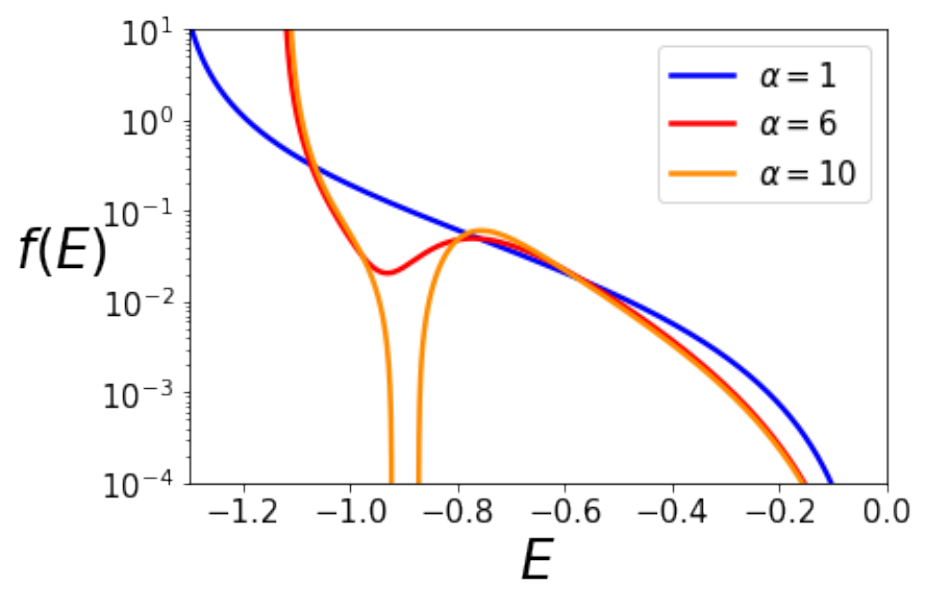}
    \caption{The isotropic distribution functions, $f(E)$, for three density profiles with 
    $\beta=5.0$, $\gamma=0.5$ and different values of $\alpha$, as indicated. As $\alpha$ increases, the system goes from stable ($\alpha=1$) to unstable ($\alpha=6$) to unphysical ($\alpha=10$). The $\alpha=6$ profile is our fiducial unstable system that is analyzed in detail in Sections~\ref{sec:evolution} and~\ref{sec:dynamics}.}
    \label{fig:examplefE}
\end{figure}

\subsection{Stability and physicality of generalized double power-law profiles}
\label{ssec:stability}

For spherical, isotropic systems, the phase-space DF $f(E)$ is determined entirely by its density profile, and can be computed using the \citet{Eddington.16} inversion:
\begin{equation}
\label{eq:eddington}
    f(E) = \frac{-1}{\sqrt{8}\pi^2} \frac{\rmd}{\rmd E} \int_0^{-E} \frac{\rmd\rho}{\rmd\Phi} \frac{\rmd\Phi}{\sqrt{\Phi-E}}\,.
\end{equation}
Here $\Phi$ is the gravitational potential and $E =\frac{1}{2}v^2 + \Phi$ is the specific energy \citep[see][]{BT2008}. Physical systems must have a positive phase-space density, i.e. $f(E)>0$ at all $E$, while systems with $\rmd f /\rmd E < 0$ at all $E$ are stable according to Antonov's theorem.

In a recent study, \citet{Baes2021} showed that for any set of isotropic $\abg$ profiles with $\gamma<\beta$, there exists an $\alpha_{\rm crit}$ such that systems with $\alpha>\alpha_{\rm crit}$ are unphysical, in that they have some range in energy where $f(E)<0$. They showed that {\it all} systems with a broken power-law density profile, which is a double power-law  with an instantaneous, non-differentiable transition, equivalent to an $\abg$ profile in the limit $\alpha \to \infty$, are unphysical.  The implication, briefly mentioned by \citet{Baes2021}, is that there must also be a range of physical $\abg$ profiles (i.e., $f(E)>0$ for all $E$) with an inflection (i.e., a range in energy where $\rmd f/\rmd E > 0$). The goals of this paper are (i) to compute for what values of $\alpha$, $\beta$ and $\gamma$ an isotropic sphere with a generalized double-power law density profile has such an inflection, thereby violating Antonov's stability criterion, and (ii) to study the evolution of such systems and characterize any ensuing instabilities.
\begin{figure*}
    \centering
    \includegraphics[width=\textwidth]{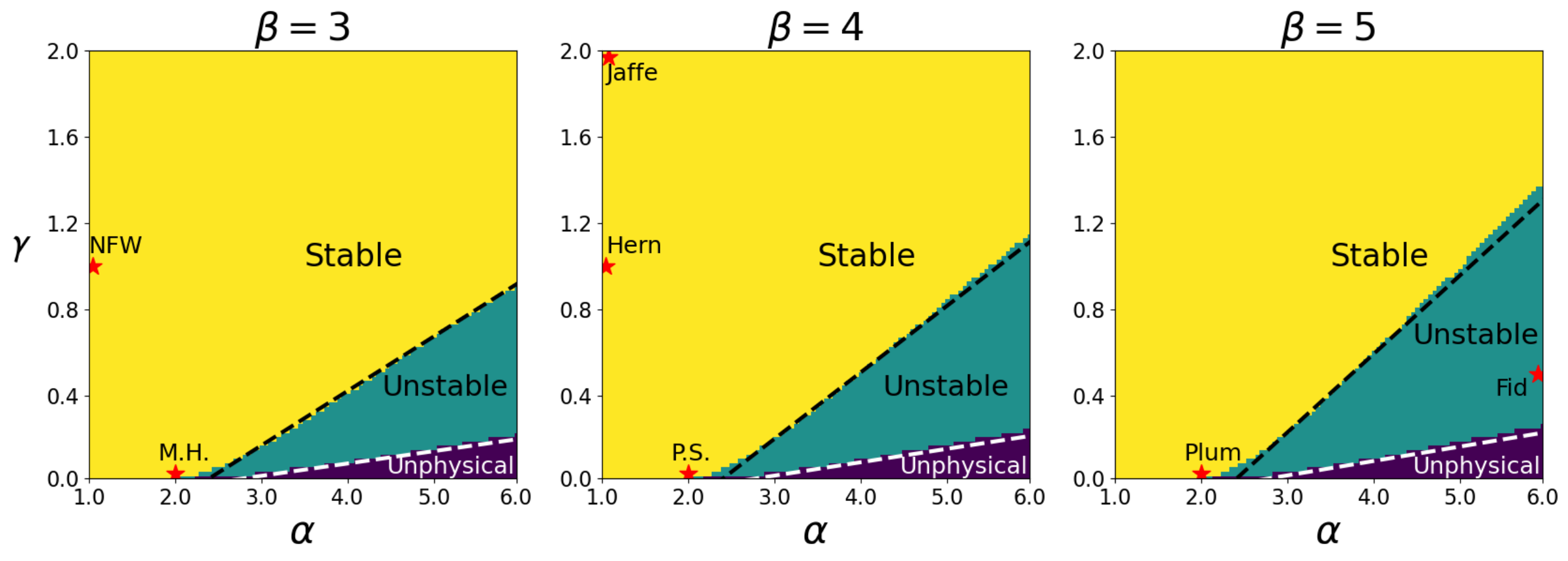}
    \caption{These figures show the regions in $\alpha-\gamma$ parameter space, for fixed $\beta$, where the system is stable ($\rmd f/\rmd E < 0$ for all $E$, shaded yellow), unstable ($\rmd f/\rmd E > 0$ at some $E$ but with $f(E) \geq 0$ at all $E$, shaded green), and unphysical ($f(E)<0$ at some $E$, shaded purple). Different panels correspond to three different values of $\beta$ as indicated. For comparison, some common density profiles are indicated, including the NFW profile \citep[][]{Navarro.etal.1997} and the modified Hubble (M.H.) profile \citep[][]{BT2008} for $\beta=3$, the perfect sphere (P.S.) profile \citep[][]{deZeeuw.85}, Hernquist (Hern) profile \citep[][]{Hernquist1990}, and Jaffe profile \citep[][]{Jaffe1983} for $\beta=4$, and the Plummer (Plum) profile \citep[][]{Plummer1911} for $\beta=5$. We also mark our fiducial unstable profile (Fid) that we focus on in this paper. In general, systems become less stable as the transition from the outer to the inner density slopes becomes sharper (larger $\alpha$). As a consequence, cored systems (small $\gamma$) are more susceptible to instability than cusped systems (large $\gamma$). The black-dashed and white-dashed lines show the fitting functions for $\gamma_{\rm stable}$ (equation~[\ref{eq:gam_stable}]) and $\gamma_{\rm physical}$ (equation~[\ref{eq:gam_physical}]) that mark the stable--unstable and unstable--unphysical boundaries, respectively.} 
    \label{fig:param_space}
\end{figure*}

We scan the $\abg$ parameter space and numerically compute the corresponding $f(E)$ using equation~(\ref{eq:eddington}). Figure~\ref{fig:examplefE} shows three examples, all of which have $\beta=5$ and $\gamma = 0.5$, but differ in their value of $\alpha$ as indicated. The DF for $\alpha=1$ (blue line) is monotonically decreasing with binding energy, $E$, which implies that $\rmd f/\rmd E < 0$ everywhere, and thus that the system is stable. When $\alpha =10$ (orange line), the distribution function has $f(E)<0$ in a small region of energy-space, rendering this model unphysical. Finally, for $\alpha=6$ (red line), the DF has an inflection, i.e., a region in energy-space where $\rmd f/\rmd E > 0$. These systems are potentially unstable and are the focal point of this study. 

Figure~\ref{fig:param_space} depicts, for three different values of $\beta$ (indicated at the top of each panel), the three regions in $(\alpha,\gamma)$ parameter space where, based on these criteria, the models are stable (yellow), unstable (green), or unphysical (purple). We emphasize that the systems labeled as ``unstable" merely violate Antonov's stability criterion and have not yet been proven to be unstable. However, as we show in this paper, all systems in this region that we have simulated are indeed found to be unstable. Hence, we refer to this region of parameter space as ``unstable'', but caution that our assessment of instability in this region is far from exhaustive. As is evident, instability {\it might} arise in systems with low $\gamma$ and/or high $\alpha$, i.e., profiles that transition sharply from a steep, outer slope to a shallow, inner slope. When this transition becomes too strong, the local minimum in the DF drops below zero and the model becomes unphysical. For comparison, we have indicated the locations of several well-known density profiles; the Hernquist profile, the Jaffe profile, the NFW profile, the Plummer sphere, the perfect sphere \citep[][]{deZeeuw.85} and the modified Hubble profile \citep[][]{BT2008}. Note that while the cuspy ones (Hernquist, Jaffe, and NFW) are safely in the stable regime, the other three cases, all of which have a constant density core with $\gamma=0$, are located very close to the stability boundary, implying that they could easily be susceptible to potential instabilities.

Based on these results, we provide linear fitting functions for the boundaries between stable, unstable, and unphysical $\abg$ profiles. For given $(\alpha,\beta)$, the minimum values of $\gamma$ that can be supported by a stable or a physical DF are:
\begin{eqnarray}
\label{eq:gam_stable}
    \gamma_{\rm stable} & = & (0.0550\beta + 0.087)\alpha - (0.136\beta + 0.181) \\
\label{eq:gam_physical}
    \gamma_{\rm physical} & = & (0.0047\beta + 0.043)\alpha - (0.013\beta + 0.102) 
\end{eqnarray}
These fitting functions for $\gamma_{\rm stable}$ and $\gamma_{\rm physical}$ are shown in Figure~\ref{fig:param_space} as the black- and white-dashed lines, respectively. We note that the fits are not perfect in that the stability and physicality boundaries are not perfectly linear. We have experimented using higher-order fitting functions, and although these improve the fit somewhat, we consider the resulting fitting functions to be unnecessarily cumbersome.


\section{Methodology}
\label{sec:method}

Having established that isotropic, spherical systems with a double power-law density profile can have an inflection in their DFs, we now examine whether such systems that violate Antonov's stability criterion are indeed unstable, and how such an instability manifests. We use numerical $N$-body simulations to evolve these systems in isolation, starting from initial conditions that are set up in equilibrium.

\subsection{Initial Conditions}
\label{ssec:ICs}

The initial conditions are established by drawing the phase space coordinates of the particles from the isotropic distribution function computed using equation~(\ref{eq:eddington}). We adopt model units for which the gravitational constant, $G$, the total mass of the system, $M$, and the scale radius, $r_\rms$, are all unity. When scaled to units corresponding to the Milky Way halo ($M = 1.37 \times 10^{12} \Msun$ and $r_\rms = 19.6 \kpc$, \citealt{McMillan2017}), the unit of time is $T \approx 35\Myr$.

Unless stated otherwise, all simulations are run using $N=5 \times 10^6$ particles. In order to increase the spatial resolution near the center of the system, we use a mass spectrum, with lower-mass particles near the center and more massive particles in the outskirts. This increases the phase-space sampling within the central region, which is our main region of interest, at no additional computational cost. Various authors in the past have devised different schemes to use such a multimass refinement. These schemes mainly differ in the physical quantity that is taken to scale with the mass of the particle. Examples used include the particle's initial radius \citep[e.g.,][]{Zemp.etal.08, Goerdt.etal.06}, its pericentric distance \citep[e.g.,][]{Sigurdsson.etal.95, Zemp.etal.08}, its specific angular momentum \citep[in the case of a disk galaxy, as in][]{Sellwood.08}, or its local phase-space density, which has been shown to minimize shot noise in the acceleration field \citep[][]{Zhang.Magorrian.08}.  Here we adopt a heuristic approach and scale the mass of particle $i$ based on its initial orbital energy, $E_i$, according to 
\begin{equation}\label{mass_spectrum}
 m_i = \frac{1}{N} \left[ 1 + \log(1+|E_i|) \right]^{-\zeta},
\end{equation}
with $\zeta$ a free parameter. Higher positive values of $\zeta$ imply a wider range of particle masses with more bound particles having a smaller mass.  Unless stated otherwise, we adopt $\zeta=15$, which results in an effective resolution near the center of the system of $N_{\rm eff} \approx 2.5 \times 10^7$, that is, an order of magnitude larger than the resolution achieved when using equal-mass particles (see Appendix~\ref{sec:appendix} for details). 

\subsection{$N$-Body Simulations}
\label{ssec:sims}

We run a set of simulations for isotropic spheres with a double power law density profile given by equation~(\ref{eq:dens}) with $\beta=5.0$ and $\gamma=0.5$, but different values for $\alpha$, ranging from $\alpha=4$ to $\alpha=8$. Based on equation~(\ref{eq:gam_stable}), systems with $\alpha \gta 3.76$ have an inflection in their DF, and therefore all the systems in our simulations violate Antonov's stability criterion. All simulations are run using the publicly available $N$-body code {\sc GyrfalcOn} \citep{Dehnen2002}. Each particle is softened using the $P_1$ softening kernel, according to which the forces between particles are computed as if they have density profiles 
\begin{equation}\label{softeningkernel}
    \rho_i(r) \propto \left[ 1+ \left(\frac{r}{\epsilon_i}\right)^2 \right]^{-7/2}\,,
\end{equation}
with $\epsilon_i$ the particle's softening length. We follow \citet{Dehnen2001} and assign each particle a softening length
\begin{equation}\label{softening}
 \epsilon_i(r) = 0.017 \left( \frac{N}{10^5} \right)^{-0.23} \left( \frac{m_i}{(1/N)} \right)^{1/3}.
\end{equation}
Hence, less massive particles, which have a larger binding energy, have a smaller softening length. As shown in Appendix~\ref{sec:appendix}, this set-up ensures that the system remains close to collisionless throughout the simulation, with no sign of mass segregation. All simulations are run using hierarchical time stepping, where the time step of each particle depends on its softening length and instantaneous acceleration, $a_i$, according to $\Delta t \sim 0.05 \sqrt{\epsilon_i/|a_i|}$. The minimum and maximum time steps are $2^{-11}$ and $2^{-6}$, respectively, which are chosen so that at most a few particles are on the smallest time step rung at any time. The total energy of the simulation is conserved to $1$ part in $\sim 10^4$.

\subsection{Calculation of the Phase-Space Distribution Function}
\label{ssec:dffromsim}

An integral part of our analysis is to study how the distribution function evolves as the system becomes unstable. To that extent, we compute the phase-space distribution of the simulation particles as a function of their instantaneous energy and angular momentum using the following method.

In a spherical system, the number of particles with energy $E$ and angular momentum $L = \vert \vect{L} \vert$ is given by
\begin{equation}\label{eq:dNdEdL}
   \frac{\rmd^2 N(E,L)}{\rmd E\,\rmd L} = f(E,L) \, g(E,L)\,,
\end{equation}
where $f(E,L)$ is the DF, and 
\begin{equation}\label{eq:dens_states}
    g(E,L) = 16 \pi^2 L \int_{r_\rmp}^{r_\rma} \frac{\rmd r}{\sqrt{2(E-\Phi)-(L/r)^2}} \ ,
\end{equation}
is the density of states. Here $r_\rmp$ and $r_\rma$ are the pericenter and apocenter distances for an orbit with energy $E$ and angular momentum $L$, which, in a spherical potential, are the roots of
\begin{equation}
 \frac{1}{r^2} + \frac{2(\Phi-E)}{L^2} = 0\,.
\end{equation}

We assume that the system remains spherical throughout its evolution, allowing us to evaluate the density of states using equation~(\ref{eq:dens_states}) and the instantaneous spherically averaged potential\footnote{Using the initial, unperturbed potential instead yields results that are indistinguishable.}. The instantaneous DF then follows from equation~(\ref{eq:dNdEdL}). As we demonstrate below, the (aspherical) unstable mode that develops saturates at a normalized amplitude of $\lesssim 10^{-2}$, justifying the assumption that the system remains spherical. In practice, we use orbital circularity $\ell=L/L_c(E)$, rather than $L$ to label the angular momentum. Here $L_\rmc(E)$ is the angular momentum of a circular orbit of energy $E$. This has the advantage that $\ell$ covers the fixed range $[0,1]$, independent of energy.

We also analyse the evolution of the one-dimensional $f(E)$, which is obtained by marginalizing $f(E,L)$ over $L$, i.e.,
\begin{align}\label{eq:marginalization}
f(E) = \dfrac{\int_0^{L_\rmc\left(E\right)} \rmd L\, f(E,L) g(E,L) }{\int_0^{L_\rmc\left(E\right)} \rmd L\, g(E,L)}.
\end{align}
For isotropic systems, the $f(E)$ thus obtained is identical to that inferred using the Eddington inversion (equation~[\ref{eq:eddington}]).

\subsection{Characterizing the instability}
\label{sec:def_triax}

In order to characterize the structural evolution of the system as it goes unstable, we use two complementary methods. The first method uses a simple expansion of spherical harmonics to quantify the power of density perturbations for modes of degree $l$ as a function of radius, while the second method uses an expansion based on biorthonormal basis functions to compute the gravitational power contained within a mode $l$ integrated over the entire system. 

Both methods presume an origin for the system of spherical coordinates. Picking such an origin is non-trivial in the presence of a dipole distortion. Using the center of mass of the entire system causes problems as a result of a few weakly bound particles with excessively large apocenters. Their discreteness can displace the center of mass enough to introduce an artificial dipole in the expansion. We therefore use the center of mass of the 50\% most bound particles as our center of expansion. We have verified that using 90\% of the most bound particles yields results that are indistinguishable. Furthermore, since {\sc GyrfalcOn} is excellent at conserving total momentum \citep[][]{Dehnen2001}, we have also experimented with expanding the density and potential around the original barycentric frame of the initial conditions, after subtracting the barycentric velocity center of mass from the velocities of all particles. This again yields results that are very similar, indicating that our results are robust to reasonable changes in how we define the center of expansion.
\begin{figure*}
    \centering
    \includegraphics[width=\textwidth]{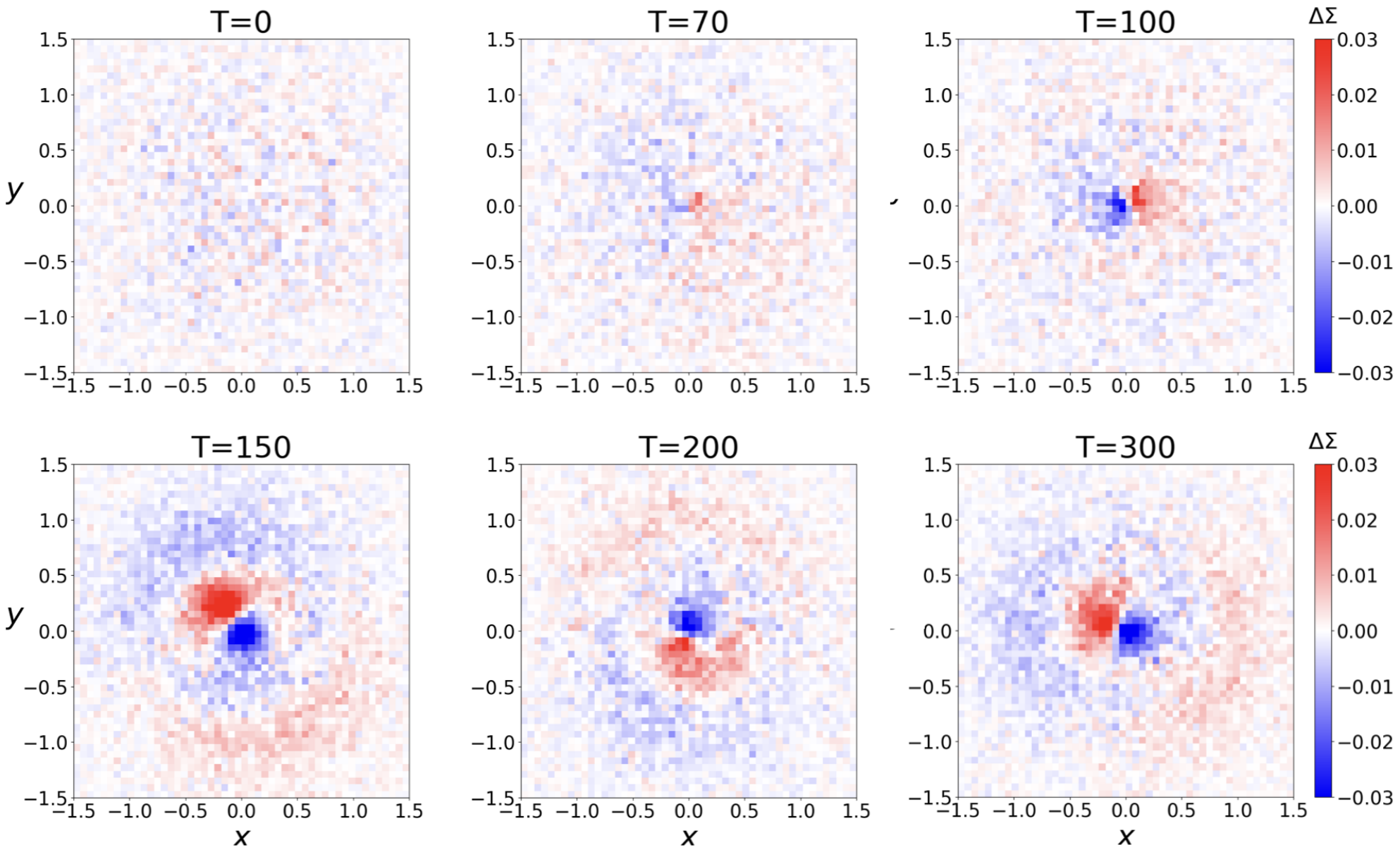}
    \caption{Projected over/underdensity ($\Delta \Sigma = \Sigma - \Sigma_0$) of our fiducial simulation
    in the $x$-$y$ plane. Different panels correspond to snapshots at different time $T$, as indicated. At $T=70$ the signatures a dipole perturpation begin to emerge, which is clearly visible by $T=100$. By $T=150$ the $l=1$ mode has saturated (see Fig.~\ref{fig:unstable_triax}), after which the perturbation continues as a long-lived, rotating soliton without any significant sign of a decline in strength.}
    \label{fig:proj_dens}
\end{figure*}

\subsubsection{Structure of the density perturbation}
\label{ssec:spherharm}

We characterize the density distribution of the system using an expansion in spherical harmonics, i.e., we define
\begin{equation}
\rho(r,\theta,\phi) = \sum_{l=0}^{\infty} \, \sum_{m=-l}^{l} a_{lm}(r) \, Y_l^m(\theta,\phi) \,.
\end{equation}
Here, $Y_l^m(\theta,\phi)$ are the orthonormal spherical harmonics, with $a_{lm}(r)$ the corresponding amplitude at radius $r$. Because $Y_l^m$ are orthogonal, the volume-weighted average amplitude of $a_{lm}(r)$ across a shell at radius $r$ is given by
\begin{equation}
\langle a_{lm} \rangle (r) =  \frac{4\pi}{V_{0,\rm shell}} 
\int_{\rm shell} \rho(r,\theta,\phi) Y_l^m(\theta,\phi) \rmd V\,,
\end{equation}
where $V_{0,\rm shell}$ is the volume of the shell. We evaluate this expression numerically using
\begin{equation}
 \langle a_{lm}\rangle(r) =  \frac{ 4\pi \sum_{\rm shell} m_i \, Y_l^m(\vect{r}_i)}{V_{0,\rm shell}}
\end{equation}
where $i$ runs over all particles in the mass shell. Finally, we define
\begin{equation}\label{eq:dens_amp}
    \langle C_l \rangle (r) = \sqrt{\sum_{m=-l}^{l} \langle a_{lm} \rangle^2(r)} \,.
\end{equation}
as a measure for the amplitude of the density distribution for a given $l$ at radius $r$.

\subsubsection{Amplitude of gravitational power}
\label{ssec:EXP}

Complementary to this straightforward expansion of the density distribution in spherical harmonics, we also calculate the gravitational power in the different spherical modes using the publicly available code EXP \citep{Petersen2022}, which uses biorthogonal basis functions to represent the density and potential of an $N$-body system. Since the systems in our paper are close to spherical, we again choose spherical harmonics as our basis functions. For each snapshot, we obtain a series of coefficients $a_{l}^m$ that represent the amplitude of the $(l,m)$ mode. The total amplitude of the $l^{\rm th}$ harmonic is then 
\begin{equation}\label{gravpower}
A_l = \sqrt{\sum_{m=-l}^l \vert a_{l}^m \vert^2} \ .
\end{equation}
Note that this amplitude corresponds to the power of harmonic $l$ in the gravitational potential energy,
\begin{equation}
W = \frac{1}{2} \int \rmd\vect{x} \, \rho(\vect{x}) \, \Phi(\vect{x})
\end{equation}
of the entire system, while the $C_l(r)$ of equation~(\ref{eq:dens_amp}) indicate the power in the density distribution of a shell at radius $r$. For more information regarding basis function expansion and EXP, we refer the reader to \citet{Petersen2022} and \citet{Petersen2025}.


\section{Results: Evolution of an unstable system}
\label{sec:evolution}

We start out the exploration of the evolution of unstable $\abg$-systems with a detailed analysis of the simulation of our fiducial system, which has $\abg=(6,5,0.5)$, placing it well within the unstable regime. Its DF is shown as the red curve in Figure~\ref{fig:examplefE}, which reveals a pronounced bump resulting in  a strong positive gradient between $E \approx -0.8$ and $E \approx -0.9$. Although we acknowledge that $\alpha=6$ is rather extreme since it creates a very sharp transition between the outer and inner slopes, it serves as a fiducial example of the dynamical behavior to be expected whenever the isotropic DF of a spherical gravitational $N$-body system has an inflection. In Section~\ref{sec:discussion} we show that lowering the value of $\alpha$ brings the system closer to the stability boundary (cf. Fig.~\ref{fig:param_space}), which reduces both the initial growth rate of the instability and the amplitude at which it saturates. Otherwise, the qualitative response of the system is largely independent of the exact value of $\alpha$. Hence, the behavior seen in our fiducial simulation seems to be representative of that for systems in the unstable region. This section focuses on the overall evolution in the structure and distribution functions of the system, while Section~\ref{sec:dynamics} presents an in-depth analysis of the orbital dynamics driving this evolution. 

\subsection{Growth of the instability}
\label{ssec:instability}

We first obtain some qualitative insight by identifying the large-scale changes in the density structure of the system as it evolves. Figure~\ref{fig:proj_dens} shows the projected over/underdensity, $\Delta \Sigma = \Sigma-\Sigma_0$, where $\Sigma_0$ is the unperturbed analytic projected density, for six key snapshots. We have projected onto the $x$-$y$ plane of the simulation volume, which happens to be close to the plane containing the dipole moment. As is evident from the snapshot at $T=0$, the discreteness of the $N$-body system gives rise to Poisson noise in $\Delta \Sigma$ of the order $\sim 10^{-2}$. Around $T \approx 70$ the first signatures of a growing perturbation at the center become evident in the projected over/underdensity. We emphasize that $T=70$ is not intrinsic to the dynamics at play, but rather simply the point in time in the simulation where the perturbation is able to be distinguished from the background noise. The perturbation primarily takes the form of a dipole, characterized by an overdensity (red) and underdensity (blue). The strength of the perturbation increases from $T=70$ to $T=150$, and remains roughly constant thereafter. Importantly, this perturbation is not stationary but rather rotating in a plane as seen in the subsequent snapshots.
\begin{figure*}
    \centering
    \includegraphics[width=\textwidth]{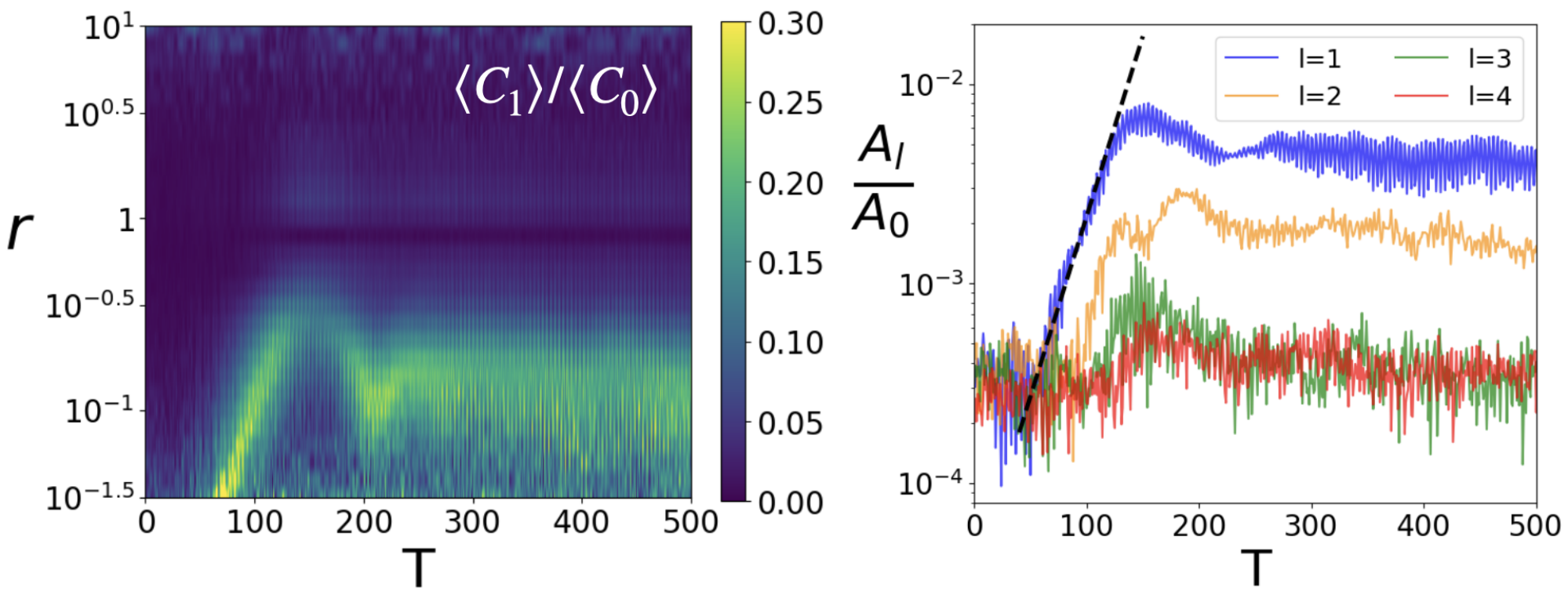}
    \caption{Left panel: the temporal and spatial evolution of the $l=1$ harmonic mode, $\langle C_1(r) \rangle$, computed using equation~(\ref{eq:dens_amp}) and normalized by the $l=0$ amplitude $\langle C_0(r) \rangle$. Time is on the $x$-axis and radius from the center of mass, $r$, is along the $y$-axis, while the color coding reflects $\langle C_1(r) \rangle/\langle C_0(r) \rangle$, as indicated in the colorbar on the right. The $l=1$ mode appears at $T\approx 70$ at the center, and the peak amplitude moves outwards to a maximum radius of $r\approx 0.4$. At late times, the peak $\langle C_1 (r) \rangle / \langle C_0(r) \rangle$ settles at $r \approx 0.1$. Note the node just below $r=1$, where the response is close to zero at all time. Right panel: the temporal evolution of the amplitude of the total gravitational power, $A_l$, computed using equation~(\ref{gravpower}) and normalized by $A_0$, for different $l$ modes, as indicated. Note how the $l=1$ mode starts out growing exponentially with time. The predicted growth rate from linear response theory (see Section~\ref{ssec:modal}) is indicated by the black dashed line, and is in close agreement with the growth rate in the simulation at early time. At $T \sim 150$, though, the $A_1/A_0$ value saturates. After a small decline in power it ultimately reaches a steady-state indicating the saturated mode is long-lived. Note also how the $A_2,A_3,A_4$ values also start growing shortly after $A_1$ grows. This apparent $l \geq 2$ power is an artifact from the imperfect representation of the $l=1$ mode.}
    \label{fig:unstable_triax}
\end{figure*}

We now apply the methods described in Section~\ref{sec:def_triax} to quantify the strength and structure of the perturbation over time. The left-hand panel of Figure~\ref{fig:unstable_triax} shows the evolution of $\langle C_1 \rangle/\langle C_0 \rangle$, computed using equation~(\ref{eq:dens_amp}), as a function of time (along the $x$-axis) and radius with respect to the system's center of mass (along the $y$-axis).  Note how the perturbation manifests itself at $T \sim 70$ as a localized maximum in $\langle C_1 \rangle/\langle C_0 \rangle$ close to the center. Subsequently, the location of this maximum moves outwards at a close to exponential rate.  At $T \sim 150$, it turns around and turns back inward, undergoes some radial oscillation, and then, by $T \sim 250$, settles down at an average radius of $r \sim 0.1$. 

The right-hand panel of Figure~\ref{fig:unstable_triax} shows the amplitude of the total gravitational power, $A_l/A_0$, in different $l$ modes, computed using EXP and normalized by the $l=0$ amplitude. The $l=1$ amplitude (shown in blue) starts growing the earliest and its initial growth is exponential with time, as predicted by linear response theory \citep{Kalnajs1977, Weinberg1991, BT2008}. 
We infer the exponential growth rate during the linear regime by fitting the slope of $\ln{\left(A_1/A_0\right)}$ as a function of time between $T=50$ and $T=100$. We find the slope to be $0.044$, which 
is in excellent agreement with the value predicted from linear response theory (see Section~\ref{ssec:modal}) and is indicated by the dashed, black line in Figure~\ref{fig:unstable_triax}. At $T\sim 150$ the $l=1$ amplitude saturates after which its power becomes roughly constant with time, except for small high-frequency modulations. As we demonstrate in Section~\ref{ssec:cusp}, these modulations reflect the radial oscillations of the dipole overdensity. 

The orange, green and red curves in the right panel of Figure~\ref{fig:unstable_triax} show that at later times, the $A_l$ for $l=2$, $3$, and $4$ also rise above the noise. However, as discussed in Section~\ref{ssec:modal}, according to linear response theory, the system is linearly unstable only to the $l=1$ mode, and not to any $l\geq 2$ modes. Therefore, the growth of this $l\geq 2$ power apparent in Figure~\ref{fig:unstable_triax} does not reflect the emergence of real $l\geq 2$ modes, but rather is an artifact that arises from the imperfect functional representation of the $l=1$ mode that is aliased into higher-order harmonics \citep{Weinberg2023}. Hence, we will not study the $l \geq 2$ signal any further.

\subsection{Motion of the cusp}
\label{ssec:cusp}

As clearly visible in Figure~\ref{fig:proj_dens}, at $T \gtrsim 70$ the system consists of an overdensity and an underdensity. By tracing the particles of this overdensity over time we establish that it corresponds to the system's original cusp (recall that the initial conditions have a central cuspy density profile with $\rho \propto r^{-0.5}$). The growing $l=1$ mode creates a net torque on the cusp, setting it in motion and dislodging it from its central location. If indeed the mode drives the motion of the cusp, it is reasonable to assume that the frequency with which the cusp circulates the center of the system is identical, or at least close to, the pattern speed, $\Omega_\rmp$, of the mode. Under this assumption, we use can the trajectory of the cusp to trace out the rotation of the $l=1$ mode, and thus infer its pattern speed. As we show in Section~\ref{ssec:modal}, this estimate is in close agreement with the pattern speed predicted using linear response theory, which justifies the assumption made. We emphasize, though, that the cusp is \textit{not} the mode itself, but rather is \textit{responding} to the mode.

The position, $\vect{r}_\rmc$, and velocity, $\vect{v}_\rmc$, of the cusp are recorded each timestep using the {\tt dens-center} tool of NEMO \citep[][]{Teuben1995}, which uses a $k$-nearest neighbor approach to determine the system's density maximum. Throughout we use $k=500$, and we have verified that none of our results is sensitive to this particular choice. We determine the orbital plane of the cusp by calculating its time-averaged angular momentum vector $\vect{L}_\rmc = \vect{r}_\rmc \times \vect{v}_\rmc$. The coordinate axes of the simulation box are then rotated so that the $z$-axis points in the direction of $\vect{L}_\rmc$, and with the origin fixed at the center of mass of the system. In this rotated coordinate system, the azimuthal angle $\phi_\rmc$ increases linearly with time to high precision, which allows us to estimate $\Omega_\rmp$ from the slope of the linear fit to $\phi_\rmc(t)$. We find $\Omega_\rmp = 0.829$ (in model units), in good agreement with the real frequency of the dipole mode (see Section~\ref{ssec:modal} below).

\begin{figure}

    \centering
    \includegraphics[width=\columnwidth]{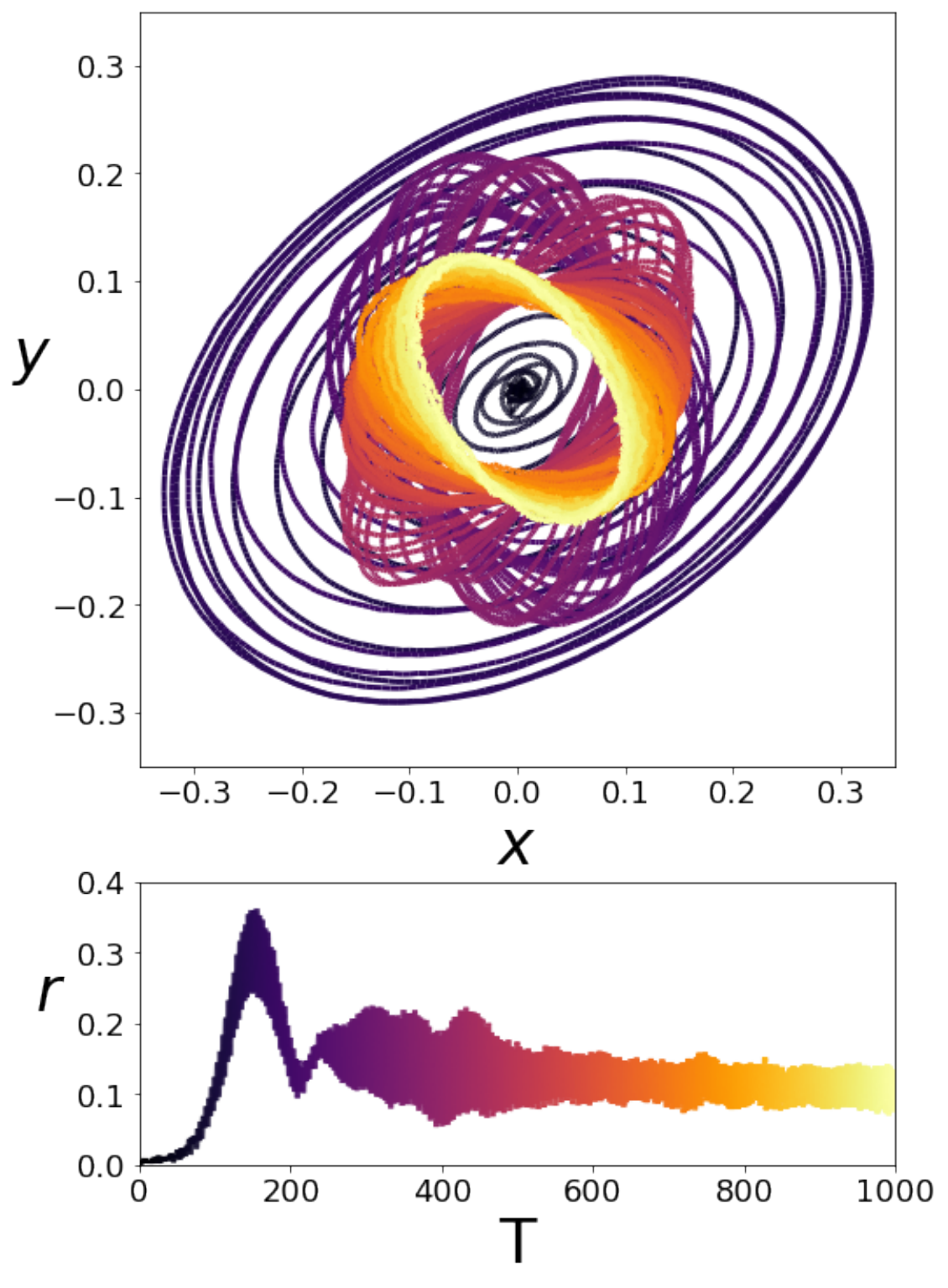}
    \caption{Top panel: the trajectory of the cusp in the $x$-$y$ plane color coded by time, with the origin fixed at the center of mass and the z-axis pointing along the cusp's angular momentum vector. Bottom panel: radial distance of the cusp from the center of mass of the system as a function of time. Note how the cusp starts at the center, moves outward as the instability grows, and eventually settles on a moderately eccentric, slowly precessing orbit.}
    \label{fig:cusp_motion}
\end{figure}

The trajectory of the cusp in its orbital plane is shown in the top panel of Figure~\ref{fig:cusp_motion}, where color is used to mark simulation time. The lower panel plots the cusp's radial distance from the origin as a function of time. During the exponential growth phase (when the response is linear), the cusp follows an outward-moving spiral away from the origin on a nearly planar orbit, in accordance with linear response theory. Once the mode goes non-linear and saturates, the cusp somewhat `sinks' inward, and its trajectory becomes more eccentric. At late times, it settles on an eccentric `orbit' at an average distance of $\sim 0.15 r_\rms$, where it continues to survive until the end of the simulation. 

The late-time trajectory of the cusp is a slowly precessing ellipse centered on the system's center of mass. The motion consists of an azimuthal frequency, given by the pattern speed $\Omega_\rmp$, and a radial oscillation with a frequency very close to $2\Omega_\rmp$. This 2:1 commensurability between the radial and azimuthal frequencies is exactly what is expected for a test particle moving in a harmonic core.  This cusp trajectory is qualitatively similar to that presented in \citet{Weinberg2023}, in which case truncation at large radii resulted in an inflection in the DF $f(E)$, which in turn triggered a dipole mode similar to that shown here. Notably, the inflection in \citet{Weinberg2023} occurred at much lower binding energy, leading to a significantly smaller pattern speed compared to our case. The late-time trajectory of the cusp in that case (see their Fig.~9) also is an ellipse, albeit significantly more radial than the one shown in Figure~\ref{fig:cusp_motion}. What exactly determines the characteristic of the trajectory of the cusp  is currently unclear and left for future study.

\subsection{Evolution of the phase-space distribution function}
\label{sec:evoloffe}

\begin{figure*}
    \centering
    \includegraphics[width=\textwidth]{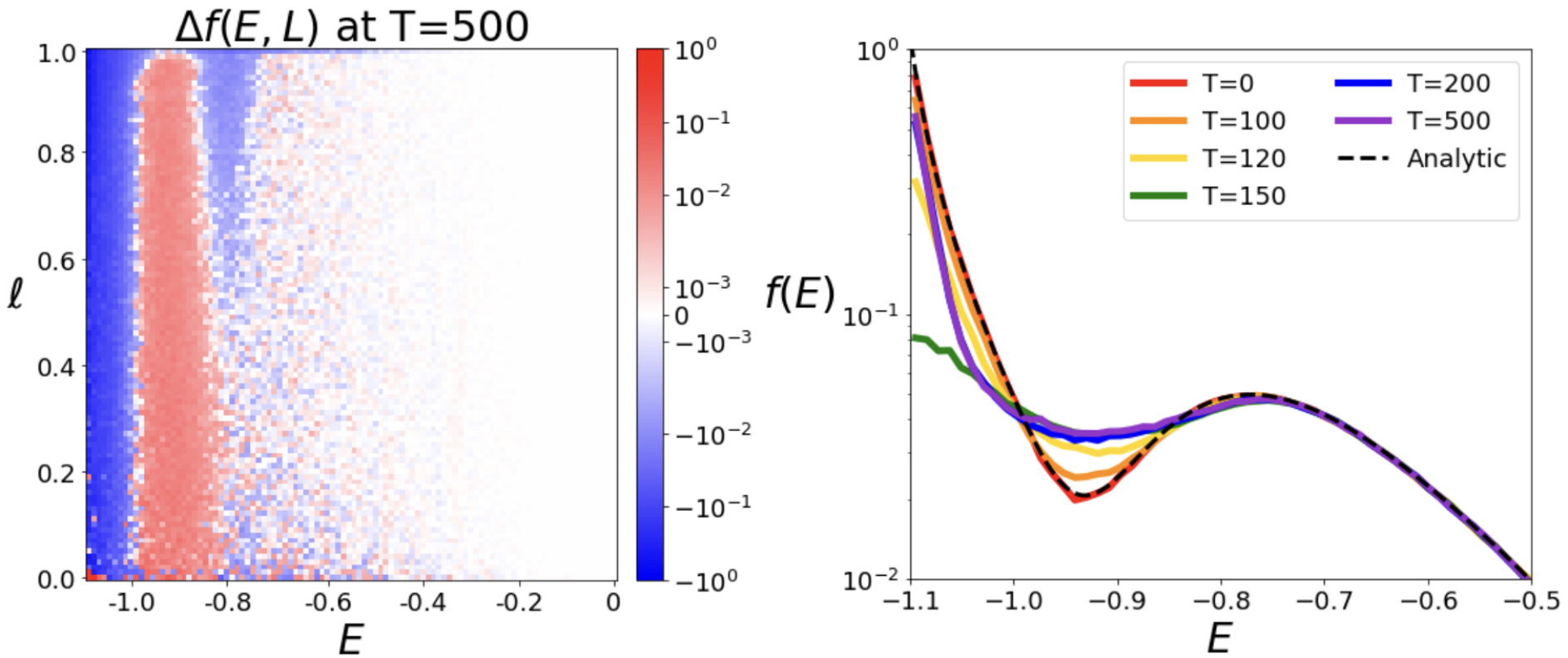}
    \caption{Left panel: the change in the two-dimensional DF, $\Delta f(E,\ell)$, between $T=500$ and $T=0$. Right panel: the evolution of $f(E)$ (equation~[\ref{eq:marginalization}]) over time. Note how the bump in $f(E)$ flattens out over time as the $l=1$ mode grows. As is evident from the left panel, this erosion of the bump is due to the fact that the phase-space density of particles with $E \gta -0.95$ increases, while that of particles with $E \lta -0.95$ decreases. As discussed in the text, this involves resonant interactions between the particles and the $l=1$ mode. Once the bump has been mostly erased, and $\partial f/\partial E \sim 0$ at the energies where the inflection used to be, the $l=1$ mode saturates, after which there is little to no further evolution in the DF.}
    \label{fig:dist_func_unstable}
\end{figure*}

At $T=0$ the system is spherical and isotropic, with an isotropic DF, $f=f(E)$, which is uniquely determined by the density distribution of the system through the Eddington inversion given by equation~(\ref{eq:eddington}). As the system becomes unstable and the $l=1$ mode grows and saturates, the DF of the system, which we calculate using the method outlined in Section~\ref{ssec:dffromsim}, undergoes changes. By following the temporal evolution of $f(E,\ell)$, we obtain valuable insight into the dynamics at play. The left panel of Figure~\ref{fig:dist_func_unstable} shows $\Delta f(E,\ell)$, defined as the change in the DF between $T=500$ and $T=0$. As is evident, the DF changes predominantly in the $E$ direction and for $E \lta -0.7$ (roughly corresponding to $r \lesssim 0.3$ in physical space). This evolution occurs in a highly structured manner, with pronounced bands in energy where the DF either increases or decreases. The right panel of Figure~\ref{fig:dist_func_unstable} shows $f(E)$, which is obtained by marginalizing $f(E,\ell)$ over $\ell$ using equation~(\ref{eq:marginalization}), with different colors corresponding to different epochs, as indicated. As is evident, over time the inflection in $f(E)$ is eroded away. When the $l=1$ power saturates at $T \approx 200$, the depression in $f(E)$ around $E \simeq -0.93$ has mostly disappeared, and from $T=200$ to $T=500$ there is virtually no further evolution in $f(E)$. Therefore, the $l=1$ power stops growing once the gradient in the DF has been mostly erased. 

Interestingly, the erosion of the bump in $f(E)$ does not run to completion, and the late-time DF, once the system has stabilized, still has a small region in $E$ where  $\partial f/\partial E$ is slightly larger than zero (see Fig.~\ref{fig:dist_func_unstable}). We speculate that this is because the $l=1$ mode introduces a slight anisotropy in the system, so the late-time DF depends on $E$ as well as $L$. The linear response matrix, after all, is a function of the three actions, not only of energy. Hence, the assumption that the one-dimensional DF in energy space fully dictates the dynamics at play breaks down. We discuss this further in Section~\ref{ssec:landau}.

\subsection{Comparison with modal analysis}
\label{ssec:modal}

Here, we compare the instability identified in our fiducial simulation to linear response theory \citep{Kalnajs.77, Fridman.Polyachenko.84, Weinberg1994, BT2008}. In short, we use a set of biorthogonal basis functions which we choose as spherical harmonics for the angular part and the solution of the Sturm-Liouville equation for the radial part \citep[see][for details]{Petersen2022} to represent the density and potential fields. We then find a simultaneous solution to the collisionless Boltzmann equation and Poisson equation. This solution can be expressed in terms of the linear response matrix $\mathbf{M}(\omega)$, given by equation~(\ref{eq:response_matrix}). The dispersion relation of the system is given by equation~(\ref{eq:dispersionrelation}), the zeros of which give the frequencies of the system's Landau (or point) modes. The imaginary part of the modal frequency is the growth/damping rate of the mode, and the real part is its pattern speed.  

The code used to calculate the response matrix is the same as in \citet{Weinberg2023} with minor updates, which implements the algorithms described in \citet{Weinberg1989, Weinberg1994}. We refer the reader to \citet{Weinberg2023} for a full description of the mathematical details and numerical implementation, as well as convergence tests. 

We identify the point modes of our fiducial $\abg=(6,5,0.5)$ system by explicitly evaluating the dispersion relation $\mathcal{D}(\omega)$ for the $l=1$ harmonic on the complex $\omega$ plane.  Specifically, we use the marching squares algorithm to find zero-valued level surfaces of the real and imaginary parts of $\mathcal{D}(\omega)$.  Their intersection determines the zeros. Since we are only interested in growing modes of the system, we restrict our calculations to $\rm{Im}(\omega)>0$.

\begin{figure}
    \centering
    \includegraphics[width=\columnwidth]{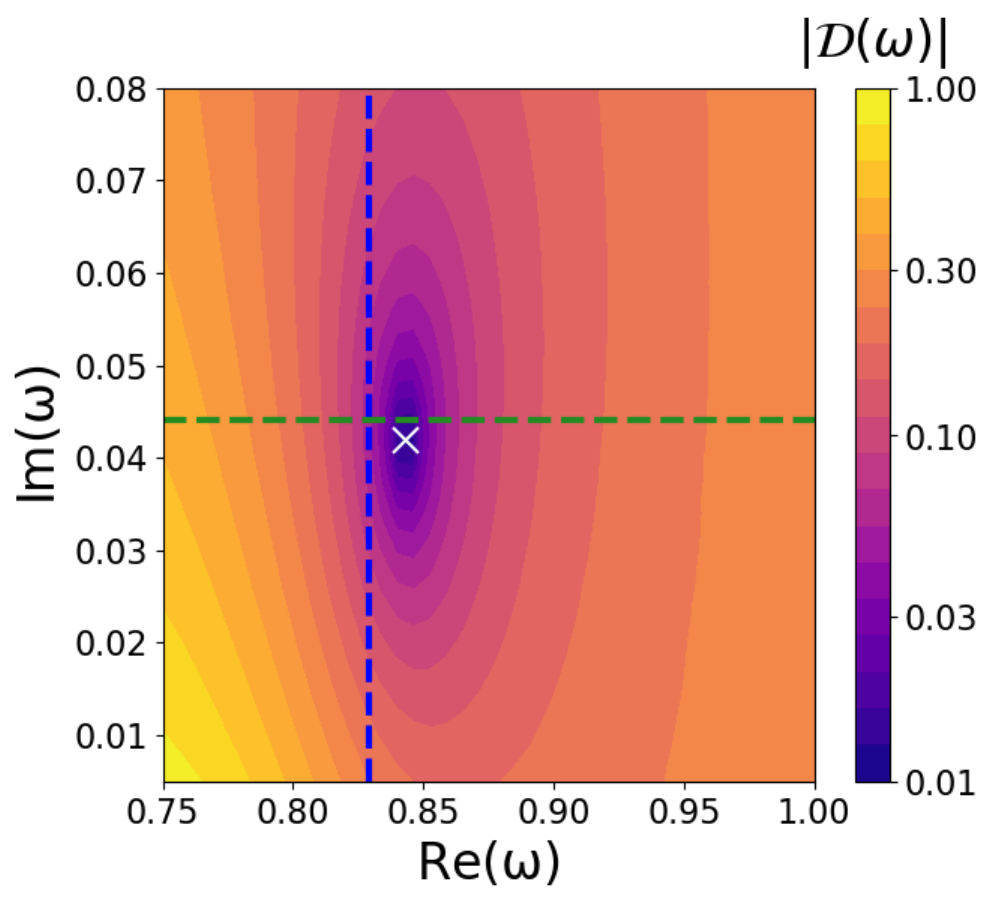}
    \caption{Contours of the dispersion relation $\left| \mathcal{D}(\omega) \right|$ of our fiducial system for the $l=1$ harmonic in the complex $\omega$ plane. The zero of the $\left| \mathcal{D}(\omega) \right|$ occurs at $\omega_{\rm mode}=(0.846,0.042)$, indicated by the white cross. 
    For comparison, the horizontal green dashed line corresponds to $\rm{Im}(\omega)=0.044$, which is the linear growth rate of the $l=1$ mode inferred from our simulation by fitting the slope of 
    $\ln{\left(A_1/A_0\right)}$ as a function of time from $T=50$ to $T=100$, which is in excellent agreement with the growth rate ${\rm Im}(\omega_{\rm mod})$ inferred from linear response theory.
    The vertical blue dashed line corresponds to $\Omega_\rmp =0.829$, which is the azimuthal frequency of the cusp, inferred from its trajectory. The fact that this is in good agreement with the frequency of the linear mode, ${\rm Re}(\omega_{\rm mod})$, indicates that the mode's pattern speed undergoes little change as it becomes non-linear.}
    \label{fig:disp_rel}
\end{figure}

Figure~\ref{fig:disp_rel} shows the two-dimensional contour of $\left| \mathcal{D}(\omega) \right|$ in the complex-$\omega$ plane. The zero of $\left| \mathcal{D} \right|$ occurs approximately at $\omega_{\rm mode}=(0.846,0.042)$, indicated by the white cross. Note the positive value of $\rm{Im}(\omega_{\rm mode})$, indicating an unstable growing mode. For comparison, the horizontal green dashed line corresponds to $\rm{Im}(\omega)=0.044$, which is the linear growth rate of the $l=1$ mode inferred from our simulation, as described in Section~\ref{ssec:instability}, and which is in excellent agreement with the value inferred here using linear response theory. The vertical blue dashed line corresponds to $\Omega_\rmp =0.829$, which is the azimuthal frequency of the cusp, inferred from its trajectory as described in Section~\ref{ssec:cusp}. This is found to be in close agreement with ${\rm Re}(\omega)$ of the $l=1$ mode. Note that the former is inferred from the response to the mode in the non-linear regime, whereas the latter is inferred from linear response theory. The fact that both are in close agreement indicates that the pattern speed of the non-linear, saturated mode remains close to its value in the linear regime. 

The unstable $l=1$ mode identified here using modal analysis only describes the perturbed system at early times, when the mode is still in the linear regime. Once the mode goes non-linear and starts to trap orbits, the response of the system is no longer adequately described by this linear mode. In fact, we have argued that the (non-linear) mode dislodges the cusp and sets it in motion. At late times, the motion of the inner halo is more intricate. The system enters a complex, quasi-periodic state in which the cusp is seen to oscillate back and forth through the center of the system. We have tried to identify the actual (non-linear, saturated) mode that we suspect to be present in addition to the dislodged cusp, but were unable to do so unambiguously. In an attempt to avoid confusion, going forward we reserve the term \textit{$l=1$ mode} for the early linear response of the system, when $A_1$ is still small and grows exponentially with time. We use the term \textit{soliton} to refer to the quasi-periodic response of the system at late times, when $A_1$ has saturated; this includes the dislodged cusp and the saturated $l=1$ mode assumed to be present as well as any other non-linear features that have originated as a result of the instability. Finally, we use \textit{cusp} to refer specifically to the set of particles that make up the original central cusp that becomes dislodged and continues to slosh back and forth as a coherent structure, always made up of the same set of particles.

\section{Orbital dynamics in an unstable system}
\label{sec:dynamics}

As shown in the previous section, the DF of the system undergoes large changes, especially in energy-space. However, the (near) constancy of the system's total energy indicates that the changes in $f(E)$ reflect a redistribution of energy among particles driven by the $l=1$ mode and its non-linear response.  In this section, we study the orbital dynamics that drive this large-scale evolution, which gives valuable insights as to the physical mechanism that leads to the growth and saturation of the $l=1$ mode.  Our analysis is performed in the rotated coordinate frame with the $z$-axis pointing along the orbital angular momentum axis of the cusp (see Section~\ref{ssec:cusp}). 

\subsection{General considerations}
\label{ssec:considerations}

In a time-independent 3D potential, each regular (non-chaotic) orbit can be defined by its canonical actions $J_i$ ($i=1,2,3$), which are conserved over time. The conjugate angles $w_i$ circulate uniformly over time, so their frequencies $\Omega_i = \dot{w_i}$ are constant. For spherically symmetric systems in $r,\theta,\phi$ coordinates, the frequencies are given by $\Omega_r$, $\Omega_\theta$, $\Omega_\phi$. Since most of the dynamics relevant for the evolution of the $l=1$ mode takes place in the $\theta=\pi/2$ plane ($z=0$), in what follows, we focus on $\Omega_r$ and $\Omega_\phi$, which are given by:
\begin{equation}\label{eq:omegar_omegaphi}
\Omega_r=\frac{2\pi}{T_r} \ , \quad \quad{\rm and}\quad \quad \Omega_\phi=\frac{2\pi}{T_\phi} \,.
\end{equation}
Here
\begin{equation}\label{eq:Tr}
T_r = 2\int_{r_\rmp}^{r_\rma} \frac{\rmd r}{\sqrt{2(E-\Phi) - L^2/r^2}} \ ,
\end{equation} 
and
\begin{equation}\label{eq:Tphi}
T_\phi=\frac{2\pi}{\Delta \phi} T_r  \quad \text{with} \quad \Delta \phi = 2L \int_{r_\rmp}^{r_\rma} \frac{\rmd r}{r^2 \sqrt{2(E-\Phi)-L^2/r^2}} \ .
\end{equation}
where $r_\rmp$ and $r_\rma$ are the pericenter and apocenter radii, respectively, for an orbit with energy $E$ and angular momentum $L$. 

In the presence of the rotating $l=1$ mode, the potential is non-spherical and time-dependent, which profoundly impacts the dynamics. The most important implication of this is the emergence of resonances between the orbital frequencies of the particles, $\Omega_r$ and $\Omega_\phi$, and the pattern speed of the $l=1$ mode, $\Omega_\rmp$. Resonances occur when the aforementioned frequencies satisfy the commensurability condition 

\begin{equation}
l_r \Omega_r + l_\phi \Omega_\phi = l_\rmp \Omega_\rmp
\label{eq:commensurability}
\end{equation}
with $\ls$ integers. In the linear limit, the actions of the perturbed orbits oscillate in time, without causing any long-term change in the distribution function. An exception are resonance orbits, for which linear theory fails to describe the response \citep{BT2008,Hamilton.Fouvry.24}. When non-linear effects such as orbit trapping are considered, close to resonances orbits can undergo permanent changes in their actions and frequencies, thus causing an irreversible evolution of the DF. As we will show in the next section, based on an orbit's proximity to a resonance, we classify orbits into three groups: non-resonant orbits, near-resonant orbits, and cusp orbits. 
\begin{figure*}
    \centering
    \includegraphics[width=\textwidth]{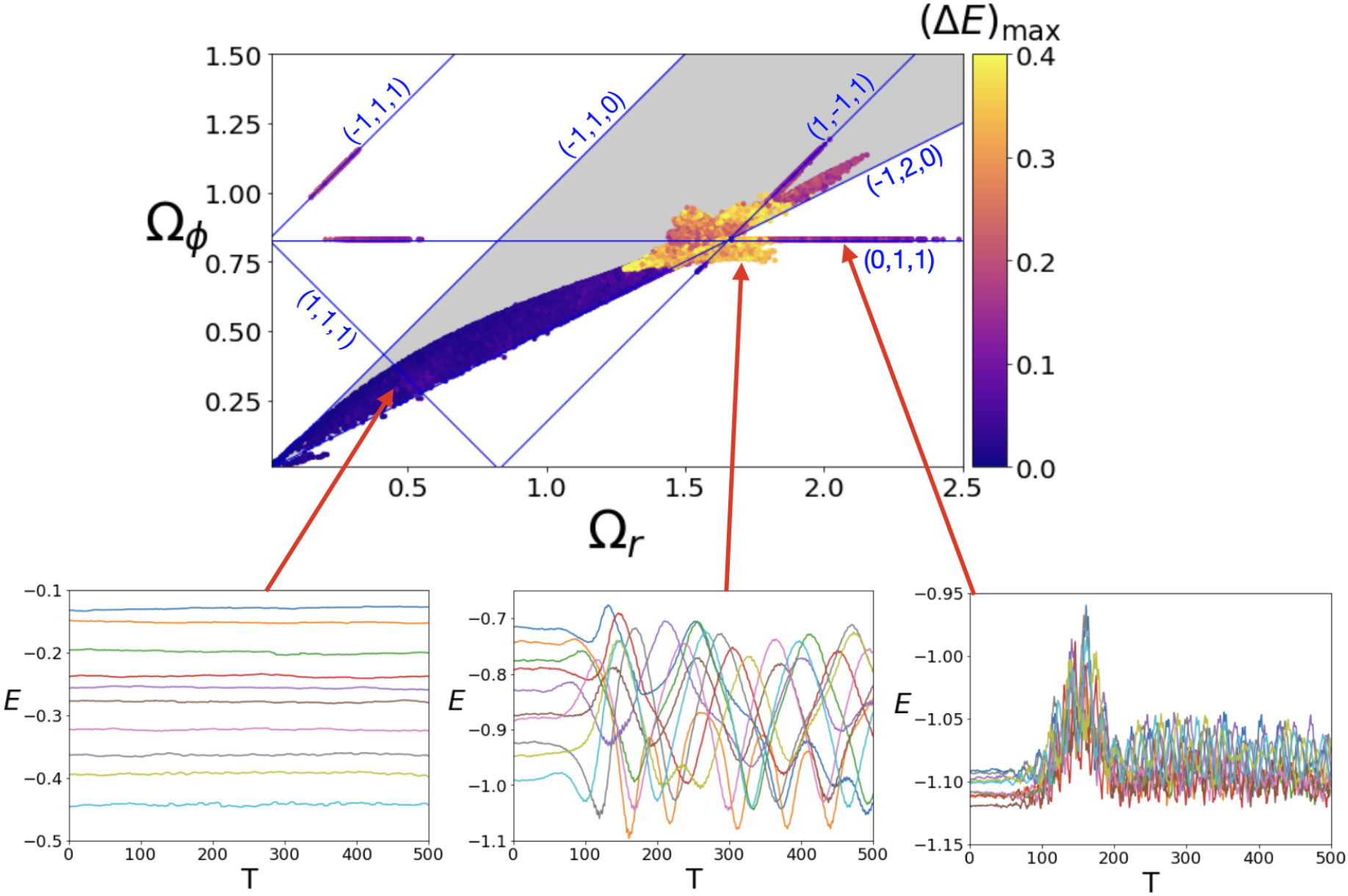}
    \caption{Top panel: frequency diagram plotting $\Omega_{\phi}$ vs. $\Omega_r$ for a random set of $10^5$ particles in the simulation. Each point is color coded by the maximum range in energy, $(\Delta E)_{\rm max} = E_{\rm max}-E_{\rm min}$, experienced by the particle over the entire duration of the simulation. Resonances are indicated by the blue lines and labeled by their $(\ls)$ values. The majority of orbits lie within $1/2 \leq \Omega_\phi/\Omega_r \leq 1$ and have low $(\Delta E)_{\rm max}$. The cusp orbits, i.e. particles that collectively make up the cusp, lie along the resonances and have moderate $(\Delta E)_{\rm max}$ values as they rapidly circulate about the cusp. The orbits that are close to (but not exactly at) resonances show the largest $(\Delta E)_{\rm max}$ values. These orbits undergo long period libration during which their energy oscillates up and down. Bottom panels: orbital energy, $E$, as a function of time for ten representative non-resonant (left), near-resonant (center), and cusp (right) orbits.}
    \label{fig:freqs}
\end{figure*}

\subsection{Frequency analysis of orbits}
\label{ssec:frequencies}

The commensurability condition (equation [\ref{eq:commensurability}]) is a linear combination of $\Omega_\phi$ and $\Omega_r$, which have simple analytic expressions for an unperturbed spherical system (equations \ref{eq:omegar_omegaphi}-\ref{eq:Tphi}). However, as the $l=1$ mode grows and changes the system's orbital configuration, the particles' $E$ and $L$ can change significantly rendering this method of computing the orbital frequencies non-trivial. Therefore, we estimate $\Omega_\phi$ and $\Omega_r$ using the Numerical Analysis of Fundamental Frequencies (NAFF) algorithm introduced by \citet{Laskar1992} and further developed by \citet{Valluri1998} and \citet{Valluri2010}. Briefly, this method expands the time series of any coordinate $q(t)$ into a Fourier time series
\begin{equation}
    q(t) = \sum_{k=1}^{k_{\rm max}} a_k \rme^{i \omega_k t}
\end{equation}
where $\omega_k$ are a set of frequencies and $a_k$ are the corresponding complex amplitudes. We use the new, python-based implementation of NAFF, called \texttt{naif}, written by \citet{Beraldo2023}.

A common method used in NAFF is to take a snapshot from the simulation, compute the gravitational potential from the particle distribution, and integrate orbits within this static potential. Since we want to properly take account of the fact that the potential is time-dependent due to the $l=1$ mode, we do not follow this approach and instead directly use the particle coordinates from the snapshot series. We verified that the snapshot cadence is high enough to obtain accurate spectra for the range of frequencies that are relevant. For the radial spectrum, the input is the complex time series $q_r(t) = r(t) + i v_r(t)$, where $r$ and $v_r$ are the radial distance and radial velocity, respectively, w.r.t. the center of mass. For the azimuthal spectrum, the input is the Poincar\'e symplectic variable $q_\phi(t) = \sqrt{2 \vert L_z(t)\vert} (\cos \phi + i \sin \phi)$, which yields a more accurate spectrum compared to using the $\phi$ variable alone \citep[][]{Papaphilippou1996, Valluri2012, Beraldo2023}. Here $\phi$ is the azimuthal angle w.r.t. the $x$-axis and $L_z$ is the angular momentum about the $z$-axis. We extract the leading frequencies of the radial and azimuthal spectra which we associate with $\Omega_r$ and $\Omega_\phi$, respectively. 
\begin{figure*}
     \centering
     \includegraphics[width=\textwidth]{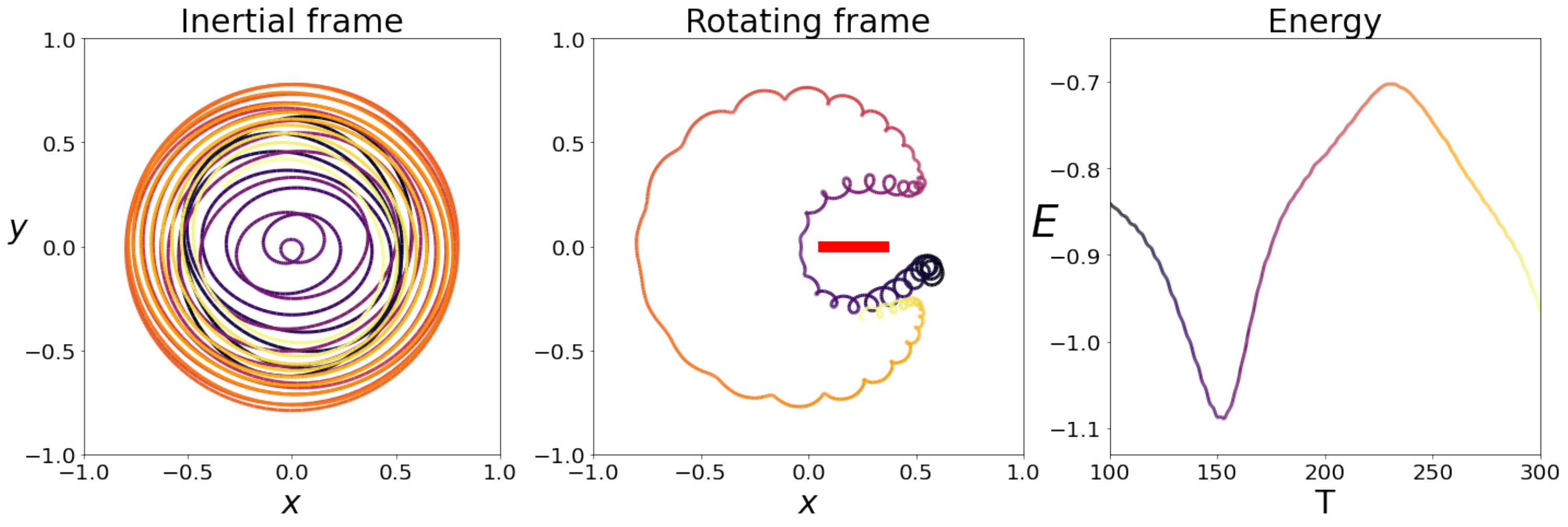}
     \caption{An example of a near-resonant orbit in the simulation. {\it Left panel}: trajectory in the inertial frame. {\it Middle panel}: trajectory in the frame co-rotating with the cusp. The red bar indicates the extent of the radial oscillations of the cusp (cf. Fig.~\ref{fig:cusp_motion}). {\it Right panel}: energy of the particle as a function of time. Note how the particle is tugged back and forth across corotation, resulting in a large-amplitude oscillations in its energy. As discussed in the text, the phase-mixing among these near-resonant orbits is responsible for eroding the bump in the DF, which in turn causes the saturation of the $l=1$ mode.}
     \label{fig:orbit}
\end{figure*}

Using the pattern speed of the soliton as inferred from the azimuthal frequency of the cusp (which is similar to the real part of the frequency of the linear mode) and the orbital frequencies of the particles, we now examine how the growth and saturation of the $l=1$ mode affects the orbital make-up, and thus the distribution function, of the $N$-body system. The top panel of Figure~\ref{fig:freqs} shows a scatter plot of $\Omega_r$ vs. $\Omega_\phi$ for a random subset of $10^5$ particles. The points are color-coded by their maximum change in energy $(\Delta E)_{\rm max} = E_{\rm max} - E_{\rm min}$, where the extrema are computed over the entire duration of the simulations. Resonances are indicated by the blue lines and labeled by their $(\ls)$ values. We identify three distinct classes of orbits in the simulation: non-resonant, near-resonant, and cusp orbits. The bottom panels plot $E$ as a function of time for 10 representative particles in each class. 
\begin{enumerate}

\item Non-resonant orbits lie inside the shaded region where $1/2 \leq \Omega_\phi/\Omega_r \leq 1$, which is the general case for rosette orbits in a spherical potential \citep[][]{BT2008}. These particles are far from resonances and have $(\Delta E)_{\rm max} \approx 0$, indicating that they do not interact with the time-dependent potential of the soliton since their energy is an integral of motion.
    
\item Near-resonant orbits, i.e. particles close to (but not exactly at) resonances, show the highest values of $(\Delta E)_{\rm max}$. They undergo long time-scale libration, which shows up as large amplitude oscillations in $E$ over time (bottom middle panel). Most of the near-resonant orbits surround the corotation resonance, and they are continuously oscillating in energy. In their non-perturbative orbit-based analysis of dynamical friction, \citet{Banik2022} identify three classes of near-corotation resonant orbits that undergo long-period libration: horseshoes, Pac-Mans, and tadpoles. We find examples of all three classes in our simulation. 
    
An example of a Pac-Man orbit\footnote{Pac-Man orbits are similar to the well-known horseshoe orbits, but while the latter circulate the $L_3$ Lagrange point, Pac-Mans, which like horseshoes and tadpoles owe their name to their appearance in the co-rotating frame, circulate the galactic center.} is shown in Figure~\ref{fig:orbit}. The left-hand panel shows the orbit in the inertial frame, and the middle panel shows the orbit in the frame corotating with the cusp, in which the cusp oscillates along the x-axis (due to its radial oscillations, see Fig.~\ref{fig:cusp_motion}), the extent of which is indicated by the red bar. The right-hand panel shows the particle's energy $E$ as a function of time. As the $l=1$ mode grows and sets the cusp in motion, the particle is trapped on a near-resonance orbit with the cusp. Initially, the particle is outside the radius of the cusp, causing it to circulate slower than the dipole. Once it comes close to the cusp, it is `pulled in' to a smaller radius, which corresponds to a decrease in its orbital energy. Now the particle circulates faster than the cusp, and the particle librates along the inner circle in the co-rotating frame. Once it comes close to the cusp again, it is `pulled outward', entering the outer circle in the co-rotating frame, which corresponds to an increase in orbital energy. Particles on horseshoe and tadpole orbits behave similarly \citep[e.g.,][]{Banik2022}. 

\item Cusp orbits, i.e. the particles that collectively make up the cusp, have moderate values of $(\Delta E)_{\rm max}$ and lie along the resonance lines in the $\Omega_r-\Omega_\phi$ plane. They are rapidly circulating about the center of the cusp, exhibiting high-frequency oscillations in $E$ as shown in the bottom-right panel. As the cusp sloshes around under the influence of the $l=1$ mode, these particles move along with it. Hence when viewed in the inertial frame, the majority of these orbits are trapped at corotation with $\Omega_\phi=\Omega_\rmp$, since the cusp is itself rotating about the center with azimuthal frequency $\Omega_\rmp$. Note that all these orbits reveal a pronounced bump in $E$ from $T \sim 70$ to $T \sim 200$, which simply reflects the radial motion of the cusp (cf. lower panel of Fig.~\ref{fig:cusp_motion}). While the bottom-right panel only shows the evolution in $E$ for a subset of particles trapped in co-rotation, the top panel shows that some particles are trapped in the $(-1,1,1)$, $(1,1,1)$, and $(1,-1,1)$ resonances, which collectively make up the $m=\pm 1$ Lindblad resonances. Although not shown, the $E$ vs. $T$ plots of particles trapped in these resonances look very similar to the particles trapped at corotation. Similar to the corotating orbits, these are all part of the original cusp that was dislodged from the center of the system. 
\end{enumerate}

\subsection{Phase-space redistribution due to resonances}
\label{ssec:phasespace}

\begin{figure}
    \centering
    \includegraphics[width=\columnwidth]{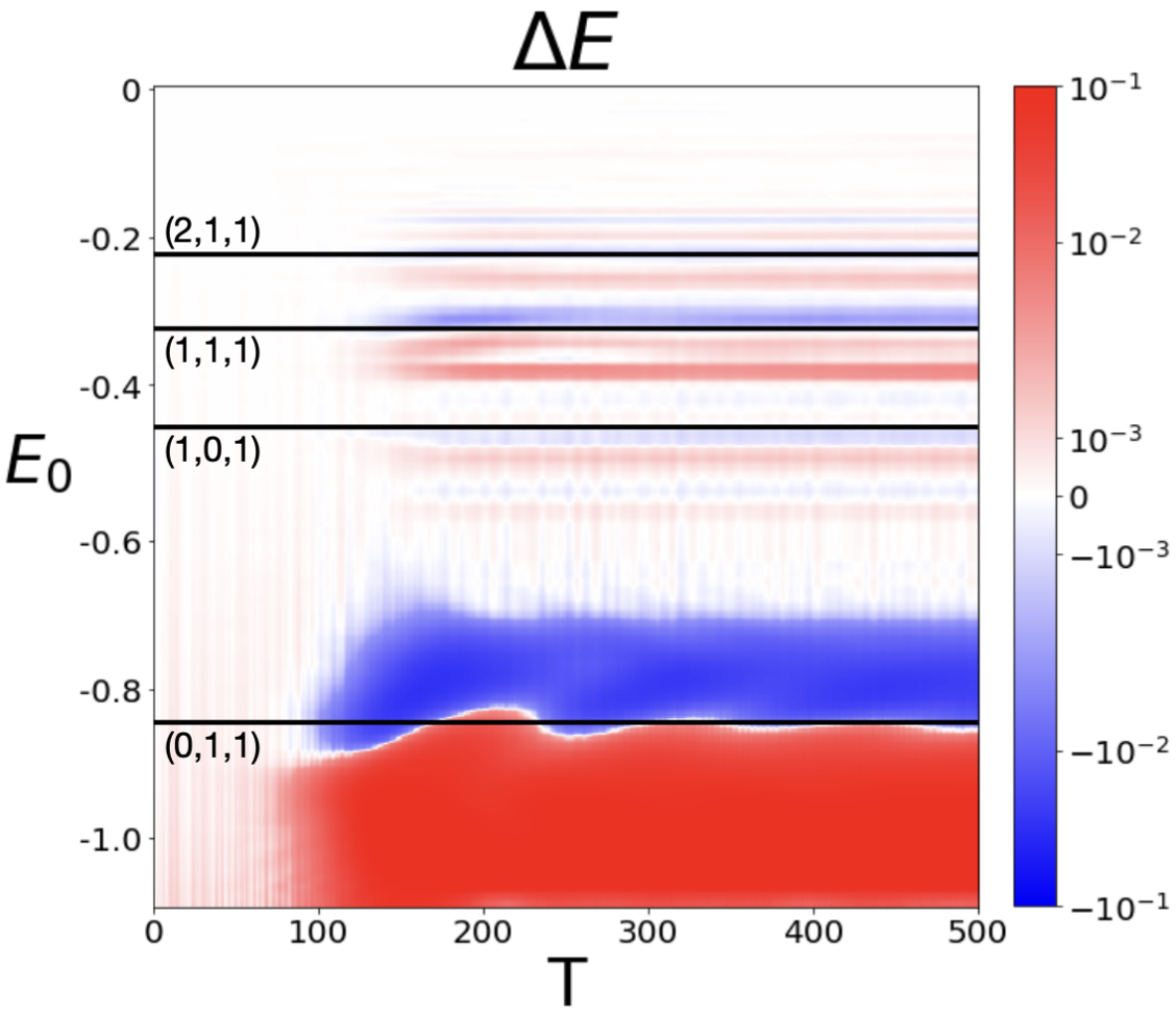}
    \caption{Map of the average fractional change in energy, $\Delta E=E(T)-E_0$ (color coded), as a function of time, $T$, and initial energy $E_0$. The average is computed at each simulation output by averaging $\Delta E$ over all particles in small bins of $E_0$. Around $T\approx 100$ when the $l=1$ starts growing, regions of large positive and negative $\Delta E$ appear for $E_0 \lta -0.7$. The positive and negative $\Delta E$ regions are separated by the $(0,1,1)$ corotation resonance. There are finer $\Delta E$ features at higher $E$ which correspond to higher order resonances in the system, marked by their $(\ls)$ values. }
    \label{fig:deltaE}
\end{figure}

To understand how the $l=1$ mode and/or the soliton impact the orbits of the particles, we study their change in energy over time. To this effect, for each particle we calculate its energy change $\Delta {E} \equiv E(T) - E_0$, and then bin the particles based on their initial energies $E_0$. Figure~\ref{fig:deltaE} shows the average $\Delta {E}$ (color coded as indicated) as a function of time and $E_0$. Around $T \sim 100$, when the $l=1$ mode starts to grow, large regions of positive and negative $\Delta {E}$ appear at $E_0 \lesssim -0.7$, with high positive and negative values of $\Delta E$. The division between the positive (red) and negative (blue) $\Delta {E}$ regions is fairly sharp and oscillatory at early times, and eventually stabilizes at $E_0 \approx -0.85$ at late times. This oscillatory behavior is due to incomplete phase mixing at early times, which results in a period of overstability, and is similar to the mechanism that can cause oscillations in the pattern speed of a bar \citep[see][]{Chiba2022}. There are also finer $\Delta {E}$ substructures at higher $E$, most of which appear as similar bands of positive and negative $\Delta {E}$.

To understand these features, we calculate $\Omega_r(E,\ell)$ and $\Omega_\phi(E, \ell)$ from equations (\ref{eq:omegar_omegaphi})-(\ref{eq:Tphi}). Both are found to have a very weak $\ell$-dependence, which allows us to compute a specific $\Omega_r$ and $\Omega_\phi$ corresponding to a particular energy, $E$\footnote{Strictly, $\Omega_r$ and $\Omega_\phi$ depend on $\ell$ as well as $E$. However the $\ell$ dependence is weak. Therefore we compute $\Omega_r(E) = \int \Omega_r (E,\ell) \rmd\ell$ and likewise for $\Omega_\phi(E)$.}. The black horizontal lines in Figure~\ref{fig:deltaE} indicate several resonances, where $l_\phi \Omega_\phi(E) + l_r \Omega_r(E) - l_\rmp \Omega_\rmp = 0$, labeled by $\ls$. As is apparent, the $(0,1,1)$ corotation resonance is the most important resonance in the system, by far. Particles that initially were located just inside of the corotation radius (lower $E$) have gained, on average, energy, while particles that were initially outside of the corotation radius (higher $E$) have predominantly lost energy. Note that the energy at the corotation resonance is $\sim -0.85$, which coincides exactly with where the gradient in the initial DF, $\partial f/\partial E$, is most positive (see right-hand panel of Fig.~\ref{fig:dist_func_unstable}). This is not a coincidence, but rather a natural outcome of the fact that the expression for the linear response matrix $\bM(\omega)$ (see equation~[\ref{eq:response_matrix}]), which governs the frequencies of the Landau modes, $\omega$, involves an integral over all actions $\bJ=(J_r, J_\phi, J_\theta)$ of the integrand $(\partial f/\partial \bJ) / (\omega - l_r \Omega_\rmr - l_\phi \Omega_\phi - l_\theta \Omega_\theta)$, which picks out the gradient in the DF where the orbital frequencies are in resonance with the Landau mode. Since $\partial f/\partial E>0$ in this energy range, there are more particles that lose energy vs. gain energy, and the response due to this imbalance leads to growth of the $l=1$ mode. We also note that at late times, the $\Delta E$ features around the corotation resonance in Figure~\ref{fig:deltaE} also reflect particle trapping by the sloshing soliton.

The finer $\Delta {E}$ features at higher $E$ correspond to higher-order resonances in the system. We do not attempt to identify all the resonances present, but three prominent ones are shown, marked by their $(\ls)$ values. Each resonance is characterized by consecutive red and blue horizontal striations, which represent particles that have lost/gained energy around that resonance. However, it is obvious that these higher-order resonances are less impactful than the corotation resonance at late times. 


\section{Discussion}
\label{sec:discussion}

We have shown that isotropic spheres with a double power-law density profile have an inflection in their DF, $f(E)$, whenever the transition from the outer density slope to the inner density slope is too rapid (i.e., $\alpha$ is too large). We have shown that this inflection causes the system to become unstable, developing a rapidly growing, rotating dipole mode. The original cusp of the system gets dislodged due to torques arising from the mode, and `orbits' along an ellipse centered on the system's center of mass, with an azimuthal frequency that remains in close agreement with the real frequency of the $l=1$ mode predicted by linear response theory. After an initial exponential growth, expected by linear theory, the mode and the dislodged cusp together saturate into a long-lived soliton that continues to slosh through the system. 

In this section, we describe the physical mechanism by which the $l=1$ mode grows and saturates, we compare this to other instabilities in both electrostatic plasmas and gravitational $N$-body systems, and we highlight the correspondence to dynamical friction. 

\subsection{Landau modes and the bump-on-tail instability}
\label{ssec:landau}

The unstable $l=1$ mode studied in this paper is a typical example of a Landau mode in a stellar dynamical system. Each of these modes describe a particular perturbation where the gravitational response of the system to the mode reproduces the mode itself. These modes can be growing or damping, depending on whether the sign of the response is positive or negative, respectively. This, in turn, depends on the shape of the DF. In particular, by distorting an otherwise stable system by slightly changing the DF in a sensitive part of phase space, one can convert a damping mode into a growing one. The $l=1$ dipole mode studied here has previously been identified in spherical King models by \citet{Weinberg1994} and \citet{Heggie2020}, but in both of these cases the $l=1$ mode was weakly damped. In our case, the mode has been rendered unstable due to an inflection in the DF, which is absent in the spherical King models. This is similar to \citet{Weinberg2023}, who found that truncating a stable double power-law distribution can introduce an inflection in the isotropic DF which renders the system unstable to a rotating dipole mode, similar to that identified here. The main difference is that in our case the dipole mode manifests in the central region due to a rapid transition in the slope of the double power-law density profile, rather than from a truncation in the outskirts. 

The instability described in this paper is analogous to the bump-on-tail (BOT) instability in plasma physics \citep[e.g.,][]{Dawson1968, Boyd.Sanderson.03}.  The spectrum of Landau modes in a homogeneous plasma is damped for a pure thermal DF. Physically, these stable Landau modes are collective electromagnetic waves.  Typically, they are excited by fluctuations, diffusively lose energy to electrons, and damp.  However, if an excess velocity \emph{bump} is close to the group velocity of the Landau mode, the wave can grow rather than damp. This increasing wave amplitude leads to diffusion of particles in velocity space that drives relaxation toward an equilibrium state, and eventually saturates the wave amplitude. These are often called micro-instabilities in the plasma literature because they are localized in space and velocity.

A stable self-gravitating $N$-body system has similar dynamics, including a spectrum of Landau-like damped modes.  For the somewhat unphysical infinite homogeneous stellar system, the dispersion relation describing the Landau modes are remarkably similar \citep[see][]{BT2008}.  For an inhomogeneous equilibrium considered here, the modes are qualitatively different and have a more intricate structure than the plasma waves, reflecting the inhomogeneous structure of the system itself.  For example, rather than matching the group velocity of the plasma wave to the velocity location of the bump, the coupling for the stellar system is dominated by the commensurabilities or resonances between the pattern and the orbital frequencies of the system (see eq.~\ref{eq:response_matrix}). A DF feature such as an inflection near any one of the resonances has the possibility of affecting the overall stability.  For an isotropic spherical system, the resonances describe nearly vertically tracks in the $(E,\ell)$ plane (Fig.~\ref{fig:unstable_triax}).  This allows us to describe the resonant dynamics through the dependence on $f=f(E)$ in the discussion that follows.

\subsection{The growth and saturation of the instability}
\label{ssec:growth}

The Landau modes of a self-gravitating system can be excited by internal perturbations (such as discreteness noise) and external perturbations (such as substructure). The net growth or damping of the perturbation depends on the gradient of the DF at each resonance. This is clear from equation (\ref{eq:response_matrix}); the response of each orbit on either side of the resonance is approximately equal but opposite in sign, and the net response depends on the phase space gradient across the resonance. Similar to the plasma BOT, the inflection in the DF changes the response of the Landau mode from damping to growing.  When the system has an inflection in its DF, one (or more) of these modes may be able to grow by changing the response to reinforcing rather than opposing the mode.  This underlies the dynamics of the dipole instability studied here; the lop-sided $l=1$ mode, which in the absence of an $f(E)$ inflection is weakly damped, grows by virtue of the inflection in the DF.

The saturation of the linear instability in a gravitational system is often modeled using the secular transport or diffusion equations pioneered by \citet{LyndenBell1972} and subsequently considered by many others \citep[e.g.,][]{Tremaine1984, Binney1988, Nelson1999, Sellwood2002, Sellwood2014, Fouvry2015}. These dynamical interactions are weakly non-linear, typically second order in the perturbation strength. The transport depends on the gradient of the DF in actions at each resonance near the commensurabilities predicted by the linear theory. This approach is often called \emph{quasi-linear theory} and assumes that the mode coupling to individual orbits is relatively weak and independent.  For our isotropic spherical stellar system, the magnitude of the diffusion depends predominantly on $\partial f/\partial E$.

In the plasma BOT context, the instability induces irreversible changes to the DF that cause evolution of the initial bump, known as plateau formation. In particular, as predicted by kinetic theory, the bump in the initial distribution function is redistributed or \emph{eroded} as the system evolves towards a detailed balance. We expect that same principle to be at work for the stellar dynamical $l=1$ dipole instability.  This picture is supported by the fact that saturation of the $l=1$ mode amplitude coincides with the time when the bump in the DF has been largely eroded away (cf. Figs.~\ref{fig:unstable_triax} and~\ref{fig:dist_func_unstable}).

At late times, the modal amplitude becomes large and explicitly non-linear effects, such as particle trapping, become important \citep[e.g.,][]{Tremaine1984, Daniel2015, Banik2022}.  Indeed, Figs.~\ref{fig:dist_func_unstable} and~\ref{fig:deltaE} show that at late times, the changes in the DF are restricted to particular bands in energy-space that straddle resonances, with corotation clearly being the dominant one.  Trapped particles librate in the frame co-rotating with the mode and together with the dislodged cusp, they give rise to a long-lived soliton that continues to rotate and slosh through the central region of the system. We emphasize that this same mechanism underlies the saturation of spiral and bar modes in a disk galaxy \citep[][]{Hamilton.24}, dynamical friction acting on a bar \citep[][]{Chiba2023, Hamilton.Fouvry.24} or a massive perturber \citep[][see also Section~\ref{ssec:conn_core_dyn}]{Banik2022}, and the fact that the Landau damping rate of an electrostatic wave in a plasma asymptotes to zero when non-linear trapping is included \citep[][]{ONeil.65}.

\subsubsection{Dependence on inflection depth}

\begin{figure}
    \centering
    \includegraphics[width=\columnwidth]{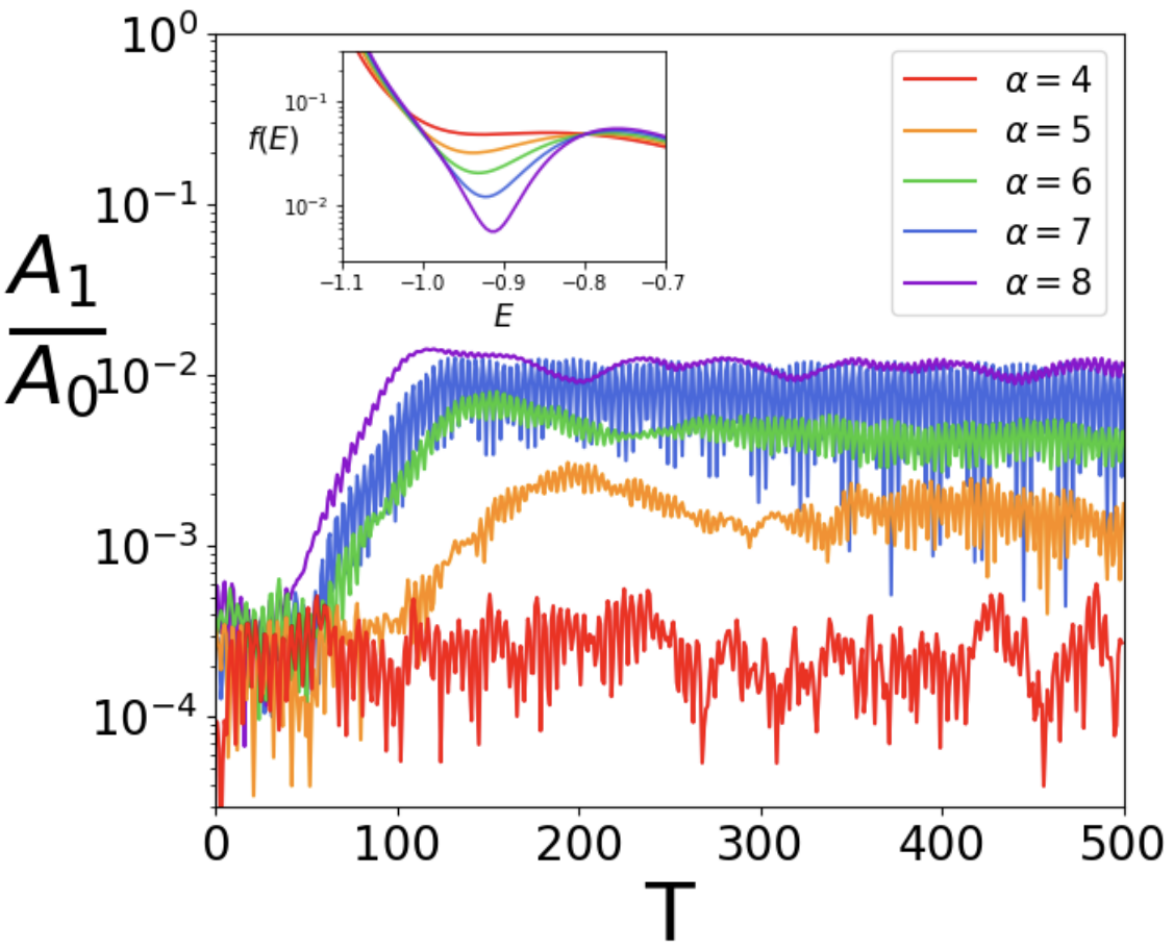}
    \caption{The normalized $l=1$ amplitude of the gravitational power, $A_1/A_0$, computed with EXP as a function of time. Results are shown for simulations of $\abg$ systems with $\beta=5.0$ and $\gamma=0.5$, but different values of $\alpha$, as indicated. The inset shows the corresponding initial DFs, $f(E)$. As $\alpha$ increases, the depth of the $f(E)$ inflection increases, making the system more unstable. This results in a higher $l=1$ saturation amplitude. Note that the system with $\alpha=4$ (in red) shows no sign of an instability. As discussed in the text, we suspect that its $l=1$ mode is too small to exponentiate above the noise.}
    \label{fig:alpha_tests}
\end{figure}

According to this picture of the growth and saturation of the $l=1$ mode, one expects that systems in which the initial inflection is more pronounced (i.e., for which the dip in $f(E)$ is deeper) develop an $l=1$ mode that saturates at a larger amplitude. After all, bump erosion will require more energy to be redistributed. To test this, we run simulations that vary $\alpha=4,5,7,8$, while keeping $\beta$ and $\gamma$ fixed at $5$ and $0.5$, respectively. As shown in the inset of Figure~\ref{fig:alpha_tests}, all these systems have an inflection in their $f(E)$, the depth of which increases with increasing $\alpha$, as the system moves further away from the stability margin (cf. Fig.~\ref{fig:param_space}). Figure~\ref{fig:alpha_tests} plots $A_1/A_0$, the total normalized $l=1$ gravitational power calculated from EXP, as a function of simulation time for the different profiles. Clearly, the saturation amplitude increases with $\alpha$, in agreement with expectations. Note also that in some cases the value of $A_1$ after saturation is rapidly oscillating with large amplitude (e.g., $\alpha=7$), while in other cases the oscillation is much weaker (e.g., $\alpha=8$). This is correlated with the peculiarities of the trajectory that the cusp ends up on; when the trajectory is highly eccentric (close to circular) $A_1$ oscillates strongly (weakly). Based on a number of numerical experiments that we have conducted, it seems that the eccentricity of the late-time trajectory of the cusp is basically random, though more work is needed to confirm this.

In the case of $\alpha=4$, the $l=1$ mode is undetectable in our simulation, even though there is a weak $f(E)$ bump. Linear stability analysis (Sec~\ref{ssec:modal}) shows that the system indeed has an unstable $l=1$ mode with a positive imaginary eigenfrequency. We therefore suspect that the system is unstable, but that the amplitude of the mode (proportional to the inverse of the imaginary part of the modal frequency) is very small. Due to the limiting resolution of the simulation, the Poisson noise is artificially high, which either prevents the mode from growing or simply makes it difficult to detect the mode above the noise. In general, as we show in Appendix~\ref{sec:appendix}, at least $\sim 10^6$ particles are required to properly resolve the $l=1$ mode in simulations. When using fewer particles, discreteness-induced relaxation can destroy the mode or artificially impact its evolution \citep[see][for a detailed discussion of the impact of collisional diffusion on the evolution of Landau modes]{Hamilton.etal.23}. Since most dark matter halos in large cosmological simulations are resolved with fewer particles, those simulations are unlikely to properly capture the potential impact of dipole instabilities on the formation and evolution of galaxies and their dark matter halos.

\subsection{Connection with dynamical friction and core dynamics}
\label{ssec:conn_core_dyn}
 
We have shown that the $l=1$ mode excites a growth promoting response owing to the overlap of the DF inflection and the corotation resonance. The resulting secular diffusion leads to energy changes of near-resonant orbits. This is characteristic of secular transport in gravitational $N$-body systems \citep[][]{Sellwood2014, Hamilton.Fouvry.24}, which underlies a variety of instabilities and associated dynamical processes such as the formation of spiral structure and the resulting radial migration of stars \citep[][]{LyndenBell1972, Sellwood2002}, bar formation \citep[][]{Sellwood.Wilkinson.93}, and dynamical friction (see below). In this subsection we focus on the latter process, pointing out the detailed correspondence with the growth and saturation of the $l=1$ mode studied here, and postulating a link to core dynamics.

As a massive perturber orbits a gravitational $N$-body system, it induces a perturbation, which then back-reacts on the perturber, resulting in dynamical friction \citep{Chandreasekhar1943}. \citet{Tremaine1984} show that according to quasi-linear theory, and in the time-asymptotic limit ($t \rightarrow \infty$), dynamical friction arises from the LBK torque \citep[][]{LyndenBell1972}, which at its core contains the gradient in the DF and the commensurability condition of equation~(\ref{eq:commensurability}). Although various authors have suggested modifications to the LBK torque formula \citep[e.g.,][]{Weinberg1989, Weinberg2004, Banik2021, Hamilton.etal.23}, they all share the key feature of secular transport being driven by resonances. Crucially, the direction of the torque, and thus the direction of the angular momentum transport, is predicted to be set by the gradient in the DF. For stable systems, $\partial f/\partial E <0$ throughout, which implies that the LBK torque is negative, resulting in angular momentum loss of the perturber causing it to sink in (i.e. friction). This is analogous to the evolution of Landau modes; if $\partial f/\partial E <0$ at the energy of the mode, then the mode will be damped. 

The LBK torque is based on quasi-linear theory and does not account for orbit trapping. Hence, it is only valid in what \citet{Tremaine1984} refer to as the ``fast regime" (i.e., a rapidly inspiraling perturber). In the slow regime, orbits are trapped at resonances, and these librating trapped orbits directly exchange energy (and angular momentum) with the perturber. This typically results in a in negative torque, i.e. friction \citep[][]{Tremaine1984, Banik2022, Chiba2022, Chiba2023}.

This picture also yields a (potentially) natural explanation for dynamical buoyancy and core stalling, which are two non-intuitive manifestations of dynamics encountered in cored galaxies that are inconsistent with predictions based on the standard treatment of dynamical friction given by \citet{Chandreasekhar1943}. Core stalling is the cessation of dynamical friction in the central constant-density core of a halo or galaxy observed in $N$-body simulations \citep[e.g.,][]{Read2006, Inoue2011, Petts2015, Petts2016}. Dynamical buoyancy is the opposite of dynamical friction, in that it makes the perturber move outward \citep{Cole2012}. Various explanations for core stalling and/or bouyancy have been advanced in the literature \citep[][]{Read2006, Petts2015, Kaur2018, Banik2021, Banik2022}, but no one has directly linked these phenomena to the gradient in the DF. For isotropic systems, the LBK torque vanishes whenever $\rmd f/\rmd E = 0$ at the orbital energy of the perturber, which should thus give rise to core stalling. Even when accounting for non-linear effects such as orbit trapping, the net torque should remain zero since a plateau in the DF implies a detailed balance between gainers and losers. This logically implies that if $\rmd f/\rmd E > 0$ (i.e., the DF has an inflection) at the orbital energy of the perturber, the net torque will be positive (enhancing), thus giving rise to dynamical buoyancy. As an analogy with plasma physics, dynamical friction and buoyancy can thus be seen as manifestations of Landau damping and inverse Landau damping, respectively; not the damping of wave energy, but rather of the orbital energy of a massive perturber.

Interestingly, dynamical buoyancy may also be a manifestation of the dipole instability. As we have shown, the growing mode dislodges the entire cusp and sets it in motion, something that is akin to buoyancy. In fact, the idea that buoyancy is related to inflections in the DF is supported by the fact that \citet{Cole2012}, who were the first to identify dynamical buoyancy in a numerical simulation, used a density profile for their host system with $\abg=(3.7,4.65,0.07)$; based on our analysis in Section~\ref{sec:abg}, such a system indeed has an inflection in its DF (i.e., $\gamma_{\rm physical} < \gamma < \gamma_{\rm stable}$), and thus would be unstable to the formation of an $l=1$ mode. We therefore suspect that the buoyancy seen in their simulation resulted from the perturber being positively torqued by the $l=1$ mode, analogous to how the cusp in our simulation was set into motion (Fig.~\ref{fig:cusp_motion}). This also explains why \citet{Meadows2020}, who studied the sinking of globular clusters in Fornax, did not find any evidence for dynamical buoyancy; the non-singular isothermal sphere that they used to model Fornax does {\it not} have an inflection in its DF. We address this potential connection between dynamical buoyancy and the $l=1$ instability in more detail in a forthcoming paper (Dattathri et al. in preparation). 

\subsection{Astrophysical Implications}
\label{ssec:obs}

If realistic galaxies and/or dark matter halos are subject to dipole instabilities, this could have a wide variety of implications. An immediate consequence of the $l=1$ mode is that it creates lopsidedness in the system, with the density peak offset from the global center of mass. If the halo that goes dipole unstable hosts a central galaxy, this is likely to manifest as an offset between the photometric center of the galaxy outskirts and its central region, or between the stellar body and the more extended HI disk. Such lopsidedness in galaxies is not uncommon, with as many as half of all late-type galaxies showing evidence for some form of lopsidedness \citep[e.g.,][]{Richter.Sancisi.94, Haynes.etal.98, Bournaud.etal.05, Reichard.etal.08, Zaritsky.etal.13}. Often the lopsidedness is evident as a strong asymmetry in the spiral structure or the location of the bar \citep[see][for a review]{Sellwood.Wilkinson.93}. A lopsided halo might also explain why the velocity field of disk galaxies is often found to be asymmetric, with large differences between the rotation curves inferred from the receding and approaching sides of the disk \citep[e.g.,][]{Jog.97, Jog.99, Swaters.etal.99}. Although all these asymmetries might potentially be the result of an $l=1$ instability, it is important to emphasize that there are various alternative ways of creating lopsidedness in disk galaxies \citep[e.g.,][]{Levine.Sparke.98, Dury.etal.08, Jog.Combes.09, Mirtadjieva.etal.11}.

The $l=1$ mode could also manifest itself as an offset between the galaxy and the halo's barycenter. Interestingly, such dark matter offsets have been detected in several external galaxies using either gravitational lensing \citep[e.g.,][]{Massey2015} or kinematics \citep[e.g.,][]{Chemin2016}. Even our own Milky Way halo has evidence of a lopsided halo \citep[][]{Saha.etal.09}. Notably, \citet{Kuhlen2013} found a dark matter offset in a zoom-in simulation of a Milky-Way analogue. Although they speculate that its origin is due to a bar-halo interaction, it is interesting that the offset of the dark matter peak only occurs after the initial cusp of the halo is transformed into a core. As we have shown in Section~\ref{sec:abg}, such a modification of the density profile could result in the creation of an inflection in the DF (if the transition in the density profile is sharp enough), indicating that the offset might instead reflect an $l=1$ instability. 

If a galaxy with a nuclear star cluster or a supermassive black hole (SMBH) experiences an instability that triggers an $l=1$ mode, it is conceivable that the star cluster or SMBH remains centered on the cusp, and thus becomes offset from the photometric center of the galaxy outskirts (as eluded to above, this basically resembles dynamical buoyancy). Interestingly, recent surveys have revealed large populations of dwarf galaxies with off-center AGN \citep[e.g.,][]{Reines2020, Mezcua2024} or off-center nuclei \citep[][]{Binggeli2000}. As these dwarf galaxies typically have extended cores in their surface brightness profiles, these offsets might well be a consequence of an $l=1$ mode, though other explanations cannot be ruled out at this point \citep[see e.g.,][]{Chowdhury2021}.

Finally, in light of our findings, it is interesting to refer to \citet{Miller.Smith.92}, who pointed out that their galaxy simulations often revealed motion of the galaxy's nucleus about the system's barycenter. After a careful study, they concluded that this was not a numerical artifact but rather the result of some instability that involved resonant interactions. Similar to the cusp in our simulations, they found the motion of their nucleus to be ``ordered and to follow reasonable trajectories". It seems likely that their simulations may have been affected by a similar $l=1$ instability as highlighted here. Note, though, that their galaxy was not modeled as an isotropic sphere with a $\abg$ density profile, but rather as a slightly flattened, rotationally supported $n=3$ polytrope. We leave it for future study to examine whether this alleged $l=1$ instability is related to some kind of inflection in the underlying DF, which must have been of the form $f=f(E,L_z)$.


\section{Summary}
\label{sec:conclusion}

We have studied the stability of self-gravitating spherical isotropic $N$-body systems with a general $\abg$ double power-law density profile. Specifically, our findings are as follows:

\begin{itemize}
    \item We have shown that whenever the density profile transitions too sharply from the outer power-law slope $\beta$ to the inner slope $\gamma$  (that is, when $\alpha$ is too large), the corresponding DF develops an inflection, resulting in a region of phase space with $\rmd f/\rmd E > 0$ (Fig.~\ref{fig:examplefE}). For a given $(\alpha,\beta)$, when $\gamma < \gamma_{\rm physical}$ the DF dips below zero, rendering the system unphysical \citep[see also][]{Baes2021}. When $\gamma_{\rm physical} < \gamma < \gamma_{\rm stable}$ the system is physical, but in violation of Antonov's stability criterion, and therefore potentially unstable (Fig.~\ref{fig:param_space}). Fitting functions for $\gamma_{\rm stable}(\alpha,\beta)$ and $\gamma_{\rm physical}(\alpha,\beta)$ are given by equations~(\ref{eq:gam_stable}) and~(\ref{eq:gam_physical}), respectively. 

    \item We have run idealized, high-resolution numerical simulations of several $\abg$ systems with an inflection in its DF, and in each case find them to be unstable to the development of a rotating dipole ($l=1$) mode. The mode initially grows exponentially with time and then saturates. The growth of the $l=1$ mode is accompanied by evolution in the DF $f(E,L)$, mostly along the $E$ axis (Fig.~\ref{fig:dist_func_unstable}). In addition, the mode dislodges the central cusp of the system from its central position, causing it to to spiral outwards with an azimuthal frequency equal to the pattern speed of the $l=1$ mode. Saturation of the mode roughly coincides with the time when the inflection has been largely erased, after which there is no further significant evolution of the DF. The cusp then settles on a trajectory that resembles an elliptical orbit, centered on the system's center of mass (Fig.~\ref{fig:cusp_motion}). 

    \item Using a detailed orbital analysis of the interaction between the $l=1$ mode and individual particles, we have demonstrated that the mode changes the energies of particles surrounding resonances. Initially, the mode is able to grow because the system's response reinforces the perturbation that causes it. Once the amplitude of the mode becomes sufficiently large, it starts to trap particles into librating orbits. This non-linear effect causes the mode to become self-confined by its own orbits, creating a soliton (analogous to the growth of an $m=2$ disk mode into a bar). The librating trapped particles undergo large periodic changes in their orbital energy (Figs.~\ref{fig:freqs} and~\ref{fig:orbit}), causing the bump in the DF to erode. Mode saturation occurs when there is a detailed balance between the particles that lose vs. gain energy, after which the DF attains a quasi-equilibrium state. 
    
    \item The instability identified using our numerical simulations is in excellent agreement with a linear mode analysis based on the matrix method of \citet{Kalnajs1977}. In particular, the mode analysis reveals that the system has an $l=1$ Landau mode (also known as a point mode), with a real frequency that is in close agreement with the pattern speed inferred from our simulation. The imaginary part of the frequency is positive, implying an unstable mode that grows over time. Although we have not formally proven this, we strongly suspect that whether this mode is unstable or damped simply depends on the gradient in the DF at the energy associated with the mode. If $\partial f/\partial E < 0$, then there are more particles that gain vs. lose energy, causing the mode to damp out (similar to Landau damping in plasmas). In contrast, when $\partial f/\partial E > 0$, the mode causes more particles close to the resonance to lose energy than gain energy, allowing it to grow. As discussed in detail in Section~\ref{ssec:landau}, the instability studied here is analogous to the bump-on-tail instability in plasma physics but for an inhomogeneous gravitational $N$-body system. 

    \item The $l=1$ mode identified here, which is associated with an inflection in the DF as a consequence of a rapid radial change in the logarithmic density gradient, is similar to the unstable $l=1$ mode in truncated NFW profiles identified by \citet{Weinberg2023} and to the weakly damped $l=1$ dipole mode in King models identified by \citet{Weinberg1994} and \citet{Heggie2020}. The picture that emerges is that most, if not all, gravitational $N$-body systems have an $l=1$ Landau mode, and whether it grows or damps is simply governed by the detailed shape of the DF. However, we emphasize that this has so far only been demonstrated for isotropic systems that have an isotropic DF. It remains to be seen whether DF inflections in more general cases, i.e. spherical systems with $f=f(E,L)$ or axisymmetric systems with $f=f(E,L_z)$, also have a similar dipole mode. 

    \item The presence of an (unstable) $l=1$ dipole mode in a galaxy or dark matter halo can have far-reaching implications, including but not limited to galaxy lopsidedness, off-center AGN and/or nuclear star clusters, and offsets between central galaxies and their halos.  The important outstanding question therefore is how common these $l=1$ instabilities are. We speculate that the various violent processes associated with the hierarchical formation of galaxies, in particular merging and core formation via explosive gas outflows driven by supernova or AGN feedback \citep[e.g.,][]{Pontzen2012, Teyssier.etal.13, Conroy2024}, will occasionally result in inflections in the DF that trigger a dipole instability. Similarly, other processes that can create cores in galaxies and/or dark matter halos, such as dynamical friction \citep[][]{ElZant2001, Goerdt2010} or core-scouring due to a binary SMBH \citep[][]{Begelman.etal.80, Quinlan1997}, may also drive the system to become unstable. Whether these processes actually result in long-lived $l=1$ solitons, and with what frequency, requires a detailed investigation using high-resolution numerical simulations. 

\end{itemize}


\section*{Acknowledgments}

We thank Chris Hamilton, Christophe Pichon, Barry T. Chiang, Kaustav Mitra, Monica Valluri, Eugene Vasiliev, and Priyamvada Natarajan for useful discussions. We thank Mike Petersen for his assistance with the EXP software package, and Leandro Beraldo e Silva for his assistance with the \texttt{naif} package. We are also grateful to Anne Lisa Vari, Jean-Baptiste Fouvry, Matthew Kunz, and Jonathan Squire for organizing the workshop ``Interconnections between the Physics of Plasmas and Self-gravitating Systems" at the KITP in Santa Barbara where part of this work was performed and where FvdB, UB, and MDW enjoyed many interactions with numerous attendees on topics relevant to this work. The KITP Santa Barbara is supported in part by the National Science Foundation under Grant No. NSF PHY-2309135. FvdB has been supported by the National Science Foundation (NSF) through grants AST-2307280 and AST-2407063. UB has been supported by the National Science Foundation Award 2209471 at the Multi-Messenger Plasma Physics Center (MPPC), and Princeton University. 


\section*{Data Availability}

Data and results underlying this article will be shared upon reasonable request to the corresponding author.


\bibliography{references}{}
\bibliographystyle{aasjournal}


\appendix

\section{Convergence and verification tests}
\label{sec:appendix}

\subsection{Verification of particle mass spectrum method}
\label{ssec:spectrum}

The mass spectrum technique described in Section~\ref{sec:method} is a powerful method to boost the central resolution while maintaining the same computational cost. Our fidicual simulation has $5 \times 10^6$ particles, implying a global average particle mass of $2 \times 10^{-7}$. However, because of the mass spectrum technique, the average particle mass, $\langle m \rangle$, inside $r = 0.1$ is $\approx 4 \times 10^{-8}$. This translates into an effective central resolution of $N_{\rm eff} \approx 2.5 \times 10^7$, a factor of $\sim 5$ higher than the resolution when using equal mass particles.
\begin{figure*}
    \centering
    \includegraphics[width=\textwidth]{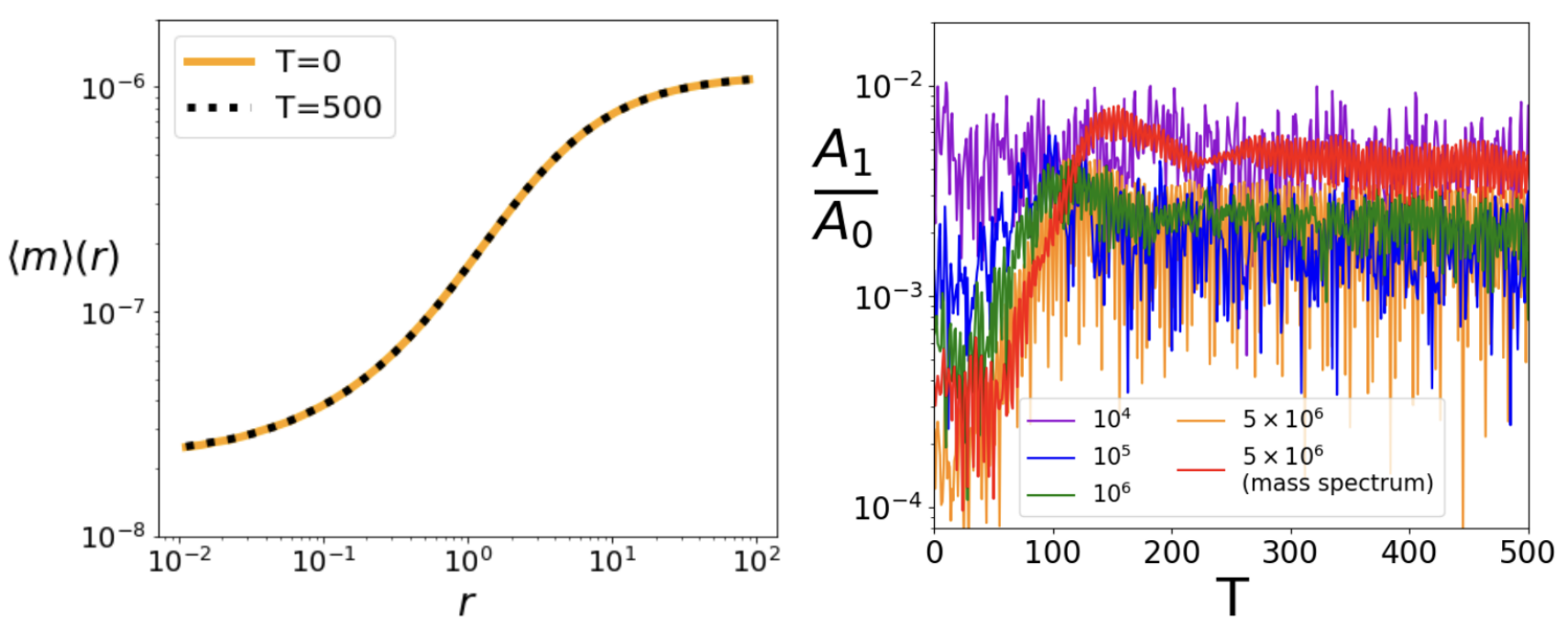}
    \caption{Left panel: the average mass of simulation particles as a function of distance from the center of mass at $T=0$ (orange) and $T=500$ (black dotted) for a simulation of a spherical system with an initial Hernquist density profile. The two lines lie exactly on top of each other, showing that the system experiences no mass segregation, despite the use of a mass spectrum. In other words, the system is collisionless, as required. Right panel: the normalized amplitude of the gravitational power of the $l=1$ mode, $A_1/A_0$, for simulation runs of our fiducial system with different resolution. Properly resolving the $l=1$ mode in the fiducial model requires of the order of $10^6$ particles. With $10^4$ particles the discreteness noise exceeds the saturation amplitude of the mode, which explains why the simulation reveals no sign of any instability.} 
    \label{fig:tests}
\end{figure*}

A potential concern of using a mass spectrum is that numerical collisionality induces mass segregation. We overcome this by scaling the softening length of the particles according to $\varepsilon \propto m^{1/3}$ (cf. equation~[\ref{softening}]), such that more massive particles have a larger softening length. In order to demonstrate that simulations run with this scheme remain perfectly collisionless, we run a simulation of a Hernquist sphere which has $\abg=(1,4,1)$ and is known to be perfectly stable (i.e., it satisfies Antonov's stability criterion). Similarly to our fiducial simulation, we use $N=5 \times 10^6$ particles whose masses are computed using equation~(\ref{mass_spectrum}) with $\zeta=15$. The orange curve in the left-hand panel of Figure~\ref{fig:tests} shows $\langle m \rangle$ as a function of radius in the initial conditions ($T=0$), which ranges from $\approx 2 \times 10^{-8}$ inside $r = 0.01$ to $\approx 10^{-6}$ at $r=100$. For comparison, the black dotted curve plots $\langle m \rangle(r)$ at $T=500$. It is indistinguishable from that at $T=0$, indicating that there is no mass segregation, as required for a collisionless system.

For completeness, we note that in our fiducial simulation discussed in the main text, which evolves a system that goes unstable, $\langle m \rangle (r)$ does slightly evolve over time. This is not due to mass segregation (collisionality), but is a real physical effect due to the radial migration of particles driven by the $l=1$ mode. As shown in Section~\ref{ssec:resolution} below, when using equal-mass particles instead, we obtain results that are qualitatively very similar (though, as expected, a bit noisier in the central region), indicating that the use of a mass spectrum does not have any adverse impact on the simulation results.

\subsection{Dependence on resolution}
\label{ssec:resolution}

The finite limiting resolution of a numerical $N$-body simulation can impact the simulation outcome in several ways. Although our simulations are to good approximation collisionless, in that there is no sign of mass segregation despite the use of a mass spectrum (see Section~\ref{ssec:spectrum} above), the unavoidable noise in the mean force field causes artificial energy diffusion that may impact the growth, decay, and saturation of Landau modes \citep[e.g.,][]{Hamilton.24}. Discreteness noise also affects the decomposition of spherical harmonics that we use to analyse the simulations (as described in Section~\ref{sec:def_triax}). In particular, it induces a ``baseline" noise level for the $l=1$ amplitude. As a consequence, one can only resolve the $l=1$ mode that arises from the inflection in the DF if its amplitude rises above this noise level. 

In order to assess how resolution impacts our results, we rerun our fiducial simulation of an isotropic sphere with $\abg=(6,5,0.5)$ with different numbers of particles and without using a mass spectrum. The right-hand panel of Figure~\ref{fig:tests} shows the temporal evolution of $A_1/A_0$ (the $l=1$ amplitude of the total gravitational power divided by the $l=0$ amplitude) for the different runs. For $10^4$ particles, there is no detectable $l=1$ mode at all, as the baseline noise level of the spherical harmonics decomposition is too high. With $10^5$ particles, there is some indication of a growing $l=1$ mode from $T=0$ to $T=100$, but it is clear that the noise level is still too high for a meaningful analysis. In addition, due to discreteness induced relaxation the amplitude of the mode quickly decays back to that of the noise-level. Properly resolving the dipole instability, with an $l=1$ mode that remains stable over time, seems to require about $10^6$ particles or more.

With a sufficient number of particles, the saturation timescale and amplitude are in rough convergence (to within a factor of $\sim 2$). We suspect that the simulation-to-simulation variance originates from noisiness in the expansion center, as well as differences in the late-time trajectory of the cusp. 



\label{lastpage}
\end{document}